\documentclass[11pt]{article}
\usepackage{epsfig}
\usepackage{amsmath}
\usepackage{amsfonts}
\usepackage{amssymb}

\numberwithin{equation}{section}

\title{Geometric B\"acklund--Darboux transformations for the KP hierarchy}
\author{G.F. Helminck\hskip 2.cm  J.W. van de Leur\footnote{JvdL
is financially supported by the
Netherlands Organization for Scientific Research (NWO).}\\
\\
\\
Faculty of Applied Mathematics,\\
University of Twente,\\
P.O.Box 217, 7500 AE  Enschede,\\
The Netherlands\\
fax: 31-53 489 4824\\
e-mail: helminck@math.utwente.nl\\
\hskip .9cm vdleur@math.utwente.nl\\
}

\newtheorem{corollary}{Corollary}[section]

\newtheorem{lemma}{Lemma}[section]
\newtheorem{proposition}{Proposition}[section]
\newtheorem{theorem}{Theorem}[section]

\begin{document}

\maketitle

\begin{tabular}[h]{lcl}
1991 MSC &:& 22E65, 22E70, 35Q53, 35Q58, 58B25\\
Keywords &:& KP hierarchy, B\"acklund-Darboux transformation,\\
&& Grassmann manifold
\end{tabular}

\begin{abstract}
In this paper it is shown that, if you have two planes in the
Segal-Wilson Grassmannian that have an intersection of finite
codimension, then the corresponding solutions of the KP hierarchy are
linked by B\"acklund-Darboux (shortly BD-)transformations. The
pseudodifferential operator that performs this transformation is shown
to be built up in a geometric way from so-called elementary
BD-transformations and is given here in a closed form. The geometric
description of elementary BD-transformations requires that one has a
geometric interpretation of the dual wavefunctions involved. This is
done here with the help of a suitable algebraic characterization of
the wavefunction. The BD-transformations also induce transformations
of the tau-function associated to a plane in the Grassmannian. They
are determined explicitly for the elementary and so-called trivial
BD-transformations. For the Gelfand-Dickey hierarchies we derive a
geometric characterization of the BD-transformations that preserves
these subsystems of the KP hierarchy. This generalizes the classical
Darboux-transformations. We also determine an explicit expression for
the squared eigenfunction potentials.

Next a connection is laid
between the KP hierarchy and the $1$-Toda lattice hierarchy. It is
shown that infinite flags in the Grassmannian yield solutions of the
latter hierarchy. These flags can be constructed by means of
BD-transformations, starting from some plane. Other applications
of these BD-transformations are a geometric way to characterize
Wronskian solutions of the $m$-vector $k$-constrained KP hierarchy and
the construction of a vast collection of orthogonal polynomials
that are parametrized among other
things by the commuting flows of the KP hierarchy. 
These systems of
polynomials play a role in matrix models.
\end{abstract}

\section{Introduction to the KP Hierarchy}

The KP hierarchy consists of a tower of nonlinear differential
equations in infinitely many variables $\{ t_n|n\ge 1\}$. It is named
after the simplest nontrivial equation in this tower,
the Kadomtsev--Petviashvili equation:
\begin{equation}
\label{KP}
\frac{3}{4}\frac{\partial^2u}{\partial t_2^2}=
\frac{\partial}{\partial t_1}\Big (\frac{\partial u}{\partial t_3}
-3u\frac{\partial u}{\partial t_1}-\frac{1}{4}
\frac{\partial^3u}{\partial t_1^3}\Big ).
\end{equation}
The solutions of these equations  belong to a ring of functions $R$
that is stable  under the operators
$\partial_n=\frac{\partial}{\partial t_n}$. The form in which one
presents these equations, is the so-called Lax form. This is an
equality between operators in the priveleged derivation
$\partial=\partial_1$ of a specific
nature.
For notational convenience we sometimes write $x$ instead of $t_1$ and
$\frac{\partial}{\partial x}$ instead of $\partial$.
The simplest way to present these equations, is to extend the
ring
$R[\partial]=\{\sum_{i=0}^n a_i\partial^i|\ a_i\in R\}$
by adding suitable integral operators to the ring. Thus it becomes
possible to take the inverse and roots of certain differential operators.
Consider the
ring $R[\partial , \partial^{-1})$ of pseudodifferential operators
with coefficients in $R$. It consists of all expressions
\[
\sum^N_{i=-\infty} a_i \partial^i\quad , \quad a_i \in R \quad
\mbox{{}for all} \;\; i ,
\]
that are added in an obvious way and multiplied according to
\[
\partial^j\circ a \partial^i
=\sum_{k=0}^\infty {j \choose k}\partial^k(a)\partial^{i+j-k}.
\]
Each operator $P=\sum
p_j \partial^j$ decomposes as $P=P_+ + P_-$ with $P_+ =
\sum\limits_{j\geq 0} p_j \partial^j$ its differential operator part
and $P_- = \sum\limits_{j<0} p_j \partial^j$ its integral operator
part. We denote by $Res_\partial P=p_{-1}$ the {\it residue } of $P$.

An operator $L\in
R[\partial,\partial^{-1})$
of the form
\begin{equation}
  \label{1.1.1}
  L=\partial + \sum_{j<0} \ell_j \partial^j\quad , \ell_j \in R\quad
  \mbox{{}for all}\;\; j<0.
\end{equation}
is called a {\it Lax operator}. We say that a Lax operator satisfies
the KP hierarchy if and only if it satisfies the system of equations
\begin{equation}
  \label{1.1.2}
  \partial_n (L) = \sum_{j<0} \partial_n (\ell_j) \partial^j =
  [(L^n)_+ , L],\quad n\geq 1,
\end{equation}
called the
{\it Lax equations} for  $L$.
These Lax equations are equivalent to the following infinite set of
equations, which are called the {\it Zakharov--Shabat equations}:
\begin{equation}
  \label{1.1.2a}
  \partial_n (L^m)_+ -\partial_m (L^n)_+=
[(L^n)_+ , (L^m)_+],\quad m,n\geq 1.
\end{equation}
Hence both (\ref{1.1.2}) and (\ref{1.1.2a}) can be seen as the
compatibility equation of the linear system
\begin{equation}
  \label{1.1.3}
  L\psi = z\psi \quad \mbox{and}\quad \partial_n (\psi) = (L^n)_+ (\psi)
\end{equation}
Notice that the equation (\ref{1.1.2}) has at least the trivial solution
$L=\partial$ and {}for this solution (\ref{1.1.3}) becomes
\[
\partial \psi = z\psi \quad \mbox{and}\quad \partial_n \psi = z^n \psi
\quad \mbox{{}for all}\;\; n\geq 1.
\]
Hence,
the function $\gamma(t) = \exp (\sum\limits_{i\geq 1} t_i z^i)$ is a
solution. The space $M$ of so-called {\it oscillating functions} is a
space {}for
which we can make sense of (\ref{1.1.3}) and that can
be seen as a collection of perturbations of this solution. It is
defined as
\[
M=\{ (\sum_{j\leq N} a_j z^j) e^{\sum t_i z^i} \mid a_i \in R,
\;\;\mbox{{}for all}\;\; i\}.
\]
Expressing $e^{\sum t_i z^i}$ in terms of the elementary Schur
functions, i.e.,
\[
e^{\sum t_i z^i}=\sum_{k=0}^\infty p_k(t)z^k
\]
one notices that the product
\[
(\sum_{j\leq N} a_j z^j)e^{\sum t_i z^i}=\sum_{\ell\in\mathbb{Z}}
(\sum_{k=0}^\infty a_{\ell-k}p_k)z^\ell
\]
is still formal. To make sense of this expression as an infinite
series in $z$ and $z^{-1}$, the coefficients
$\sum_{k=0}^\infty a_{\ell-k}p_k$ for all $\ell\in\mathbb{Z}$ have to be
well defined functions of $t=(t_n)$. In that light it is natural to
have a context such that for each element $t$
\[
\sum_{j\leq N} |a_j(t)|^2<\infty
\quad\mbox{and}\quad
\sum_{k=0}^\infty |p_k(t)|^2<\infty.
\]
Such a context has been given in \cite{SW} and will be considered in
section 2.

The space $M$ becomes a $R[\partial, \partial^{-1})$-module by the
natural extension of the actions
\[
\begin{array}[h]{lcl}
b\{(\sum_j a_j z^j) e^{\sum t_i z^i}\} &=& (\sum_j ba_j z^j) e^{\sum t_i
  z^i}\\
\partial\{ (\sum_j a_j z^j) e^{\sum t_i z^i}\} &=& (\sum_j \partial(a_j)
z^j + \sum_j a_j z^{j+1} ) e^{\sum t_i z^i} .
\end{array}
\]
It is even a free $R[\partial , \partial^{-1})$-module, since we have
\[
(\sum p_j \partial^j) e^{\sum t_i z^i} = (\sum p_j z^j )e^{\sum t_i
  z^i} .
\]

An element $\psi$ in $M$ is called  an {\it oscillating function of type}
$z^{\ell}$,
if it has the
form
\begin{equation}
\label{1.6h}
\psi (z) = \psi(t,z)=\{ z^{\ell} + \sum_{j<\ell} \alpha_j z^j\} e^{\sum t_i
z^i} .
\end{equation}
The fact that $M$ is a free $R[\partial, \partial^{-1})$-module,
permits you to show that each oscillating function of type $z^{\ell}$ that
satisfies (\ref{1.1.3}) gives you a solution of (\ref{1.1.2}). This
function is then called a {\it wavefunction} of the KP-hierarchy.
Notice that we can write
$\psi(z)=P(\partial)e^{\sum t_i z^i}$. It
is not difficult to see that the equations (\ref{1.1.3}) for
$L=P\partial P^{-1}$ are
equivalent to the Sato--Wilson equations for $P$:
\begin{equation}
  \label{1.3.1}
   \partial_n( P) P^{-1} = - (P
   \partial^n P^{-1})_-.
\end{equation}

Besides the space of oscillating functions $M$ it is also convenient
to have at one's disposal its ``adjoint'' space $M^*$ consisting of
all formal products
\[
\{\sum_{j\le N}a_jz^j\} e^{-\sum t_i z^i}, \quad\mbox{with } a_j\in
R\ \mbox{for all }j.
\]
The ring $R[\partial,\partial^{-1})$ acts as expected on $M^*$ by the
natural extension of the actions
\[
\begin{array}[h]{lcl}
b\{(\sum_j a_j z^j) e^{-\sum t_i z^i}\} &=& (\sum_j ba_j z^j) e^{-\sum t_i
  z^i}\\
\partial\{ (\sum_j a_j z^j) e^{-\sum t_i z^i}\} &=& (\sum_j \partial(a_j)
z^j - \sum_j a_j z^{j+1} ) e^{-\sum t_i z^i} .
\end{array}
\]
This is also a a free $R[\partial , \partial^{-1})$-module, since we have
\[
(\sum p_j (-\partial)^j) e^{-\sum t_i z^i} = (\sum p_j z^j )e^{-\sum t_i
  z^i} .
\]
There is a bilinear pairing $\mathfrak{R}:M\times M^*\to R$
defined as follows: if $\phi=\phi(t,z)=(\sum_j a_j(t) z^j)e^{\sum t_i
z^i}$
belongs to $M$ and $\psi=\psi(t,z)=(\sum_k b_k(t)z^k)e^{-\sum t_i
z^i}$
is an element of $M^*$, then we put
\[
\begin{aligned}
\mathfrak{R}(\phi,\psi):&=Res_{z=0}\{(\sum_j a_j(t) z^j)(\sum_k
b_k(t)z^k)\}\\
\ &=\sum_{k\in\mathbb{Z}}a_{-k-1}(t)b_k(t),
\end{aligned}
\]
which is clearly a finite sum and hence belongs to $R$.

On $R[\partial , \partial^{-1})$ we have an anti-algebra
morphism called {\it taking the adjoint}. The adjoint of $P=\sum p_i
\partial^i$ is given by
\[
\begin{aligned}
P^* =& \sum_i (-\partial)^i p_i =\sum_i (-)^i\sum_{k=0}^\infty
{{i}\choose{k}} \partial^k(p_i)\partial^{i-k}\\
=&\sum_\ell\left \{\sum_{k=0}^\infty(-)^{\ell+k}{{\ell+k}\choose{k}}
\partial^k(p_{\ell+k})\right \}\partial^\ell .
\end{aligned}
\]
There is an important connection between the bilinear form $\mathfrak{R}$
and taking the adjoint.
\begin{theorem}
\label{t1.1h}
Let $\phi(t,z)=P(t,\partial)e^{\sum t_i z^i}\in M$ and
$\psi(t,z)=Q(t,\partial)
e^{-\sum t_i z^i}\in M^*$, then
\begin{equation}
\label{bil0}
\begin{aligned}
(P(t,\partial)Q(t,\partial)^*)_-
&=\mathfrak {R}(\phi(t,z),\partial^{-1}\circ\psi(t,z))\\
&:=-\sum_{n=0}^\infty
\mathfrak {R}(\phi(t,z),\frac{\partial^n\psi(t,z)}{\partial
x^n})(-\partial)^{-n-1}.
\end{aligned}
\end{equation}
\end{theorem}
{\bf Proof.} The proof consists of showing, by a direct computation,
that the coefficients of
$\partial^{-n-1}$ on both sides of (\ref{bil0}) are equal.
 Let $P=\sum_j a_j\partial^j$ and
let $Q=\sum_k b_k\partial^k$. First we compute the coefficient
$\mathfrak {R}(\phi(t,z),\frac{\partial^n\psi(t,z)}{\partial x^n})$.
Since
\[
\begin{aligned}
\partial^n(\psi)=&\left \{ \sum_{\ell=0}^n {n\choose \ell} \sum_k
\partial^{n-\ell}(b_k)(-z)^{k+\ell}
\right \}e^{-\sum t_i z^i}\\
=&\left \{ \sum_p \left (\sum_{\ell=0}^n {n\choose \ell} (-)^p
\partial^{n-\ell}(b_{p-\ell})\right )z^p
\right \}e^{-\sum t_i z^i},
\end{aligned}
\]
we get
that
\[
\mathfrak {R}(\phi(t,z),\frac{\partial^n\psi(t,z)}{\partial x^n})
=\sum_p a_{-p-1}\sum_{\ell=0}^n {n\choose \ell} (-)^p
\partial^{n-\ell}(b_{p-\ell}).
\]
On the other hand we have
\[
\begin{aligned}
PQ^*=&\sum_{j,k} a_j\partial^j(-\partial)^kb_k=\sum_{j,k}
(-)^ka_j\partial^{j+k}b_k\\
=&\sum_{j,k}
(-)^ka_j\sum_{r=0}^\infty {{j+k}\choose
r}\partial^r(b_k)\partial^{j+k-r}\\
=&\sum_s\left \{
\sum_j\sum_{r=0}^\infty {{r+s}\choose
r} (-)^{r+s+j}a_j\partial^r(b_{r+s-j})
\right \}\partial^s.
\end{aligned}
\]
Hence the coefficient of $\partial^{-n-1},\ n\ge 0$, equals
\[
\begin{aligned}
(-)^{n+1}\sum_j\sum_{r=0}^\infty {{r-n-1}\choose
r}& (-)^{r+j}a_j\partial^r(b_{r-j-n-1})=\\
=&(-)^{n+1}\sum_j\sum_{r=0}^n {{n}\choose
r} (-)^{j}a_j\partial^r(b_{r-j-n-1})\\
=&(-)^{n+1}\sum_j(-)^ja_j\sum_{\ell=0}^n{{n}\choose
\ell}\partial^{n-\ell}(b_{-\ell-j-1})\\
=&(-)^n\sum_{p}(-)^p a_{-p-1}\sum_{\ell=0}^n{{n}\choose
\ell}\partial^{n-\ell}(b_{p-\ell})\\
=&(-)^n\mathfrak{R}(\phi(t,z),\frac{\partial^n\psi(t,z)}{\partial x^n}).
\end{aligned}
\]
This proves the claim in the theorem.\hfill$\square$\

\noindent
If $\phi=P(t,\partial)e^{\sum t_i z^i}$ is an oscillating function of type
$z^\ell$, then we call
$\phi^*=P(t,\partial)^*e^{-\sum t_i z^i}$ in $M^*$ the adjoint of
$\phi$. It is a dual oscillating function of type $z^{-\ell}$. The
foregoing theorem shows that for all $n\ge 0$
\begin{equation}
\label{1.8h}
\mathfrak{R}(\phi(t,z),\frac{\partial^n\phi^*(t,z)}{\partial t_1^n})=0.
\end{equation}
This property even characterizes $\phi^*$ among the dual oscillating
functions of type $z^{-\ell}$. For, if $\psi=Q(t,\partial)
e^{-\sum t_i z^i}\in M^*$ is such a function satisfying (\ref{1.8h})
with $\phi^*$ replaced by $\psi$, then we have according to the
theorem
\[
PQ^*=\partial^0+(PQ^*)_-=1.
\]
In other words, $Q=(P^*)^{-1}$ and $\psi=\phi^*$. We will use this
criterion later on and therefore  we resume it in a
\begin{lemma}
\label{l1h}
Let $\phi$ be an oscillating function of type $z^\ell$ and $\psi$ a
dual oscillating function of type $z^{-\ell}$. Then $\psi$ is the
adjoint of $\phi$ if and only if it satisfies
\[
\mathfrak{R}(\phi(t,z),\frac{\partial^n\psi(t,z)}{\partial t_1^n})=0\quad
\mbox{for all }n\ge 0.
\]
\end{lemma}
If $\psi\in M$ is a wavefunction for the KP hierarchy, then its
adjoint $\psi^*$ satisfies a similar set of linear equations, viz.,
\begin{equation}
\label{1}
L^*\psi^* = z\psi^* \quad \mbox{and}\quad \partial_n (\psi^*) = -(L^n)_+^*
(\psi^*),
\end{equation}
where $L^*=(P\partial P^{-1})^*=(P^*)^{-1}(-\partial)P^*$. The first
of these equations is a direct consequence of the definition for
\[
(P^*)^{-1}(-\partial)P^*(P^*)^{-1}e^{-\sum t_i
z^i}=(P^*)^{-1}(-\partial)
e^{-\sum t_i z^i}=z(P^*)^{-1}e^{-\sum t_i z^i}=z\psi^*.
\]
The second follows from the Sato--Wilson equations for $P$. By taking
the adjoint of relation (\ref{1.3.1}) we get
\[
(\partial_n( P) P^{-1})^*= P^{*-1}\partial_n(P^*)= - (P
   \partial^n P^{-1})^*_-=-(P^{*-1}(-\partial)^nP^*)_-=-((L^*)^n)_-.
\]
Since $\partial_n(P^{*-1})=-P^{*-1}\partial^n(P^*)P^{*-1}$, these
equations combine to give
\begin{equation}
\label{1.10h}
\partial_n(P^{*-1})P^*=((L^*)^n)_-.
\end{equation}
As we have that
\[
\begin{aligned}
\partial_n(\psi^*)&=\partial_n(P^{*-1}e^{-\sum t_i
z^i})=\partial_n(P^{*-1})e^{-\sum t_i
z^i}-z^nP^{*-1} e^{-\sum t_i
z^i}\\
\ &=((L^*)^n)_-\psi^*- (L^*)^n\psi^*=-((L^*)^n)_+\psi^*,
\end{aligned}
\]
which is exactly the second equation of (\ref{1}). Reversely, if the
adjoint of an oscillating function of type $z^\ell$ satisfies the
equations in (\ref{1}), then $\psi$ is a wavefunction of the KP
hierarchy. The equations in (\ref{1}) namely imply equation
(\ref{1.10h}) and by taking the adjoint of it, we get the Sato--Wilson
equations for $P$.

In the present context, where the splitting of the oscillating
functions in the exponential factor $e^{\sum t_i
z^i}$ and a factor that has at most a pole at infinity is a formal
affair. One also does not have at one's disposition the famous
bilinear identity
(see \cite{DJKM}):
\begin{equation}
\label{bil}
Res_{z=0} \psi(t,z)\psi^*(s,z)=0,
\end{equation}
to characterize the oscillating functions $\psi$ of type $z^\ell$ that
are wavefunctions  for the KP hierarchy.
We will see in the next section that the relation (\ref{bil}) makes
sense in the convergent setting of Segal and Wilson.

Another item in
the theory
 of the KP hierarchy is the tau-function $\tau(t):=\tau((t_i))$. It was
introduced by the Kyoto school, see \cite{DJKM}, and relates to the
wavefunction $\psi$ and its adjoint wavefunction $\psi^*$ by the
formula's
\begin{equation}
\label{D1}
\psi(t,z)=z^{\ell}\frac{\tau(t-[z^{-1}])}{\tau(t)}e^{\sum t_iz^i}
\quad\mbox{and }
\psi^*(t,z)=z^{-\ell}\frac{\tau(t+[z^{-1}])}{\tau(t)}e^{-\sum t_iz^i},
\end{equation}
where $[\lambda]
=(\lambda,\frac{1}{2}\lambda^2,
\frac{1}{3}\lambda^3,\frac{1}{4}\lambda^4,\ldots)$.
In the general set-up of the oscillating functions its existence is by
no means clear,
but in the Segal--Wilson context such a $\tau$ exists. In order that
the right-hand sides of the equations (\ref{D1}) are in $M$ resp.
$M^*$, there should hold the Taylor formula for $\tau$. By applying
this to (\ref{D1}) we get for all first order derivatives of $\tau$
\begin{equation}
\label{tau}
\frac{\partial_k(\tau)}{\tau}=Res_{z=0}z^k\left ((\frac{\partial}{\partial z}
-\sum_{j=1}^{k-1}z^{-j-1}
\partial_j)\log (z^{-\ell}e^{-\sum t_iz^i}\psi(t,z))\right ).
\end{equation}
{}From this formula we see that if there are 2 tau-functions satisfying
(\ref{D1}),
then their quotient is independent of the variables $(t_i)$. If a
wavefunction $\psi$ of the KP hierarchy posesses a function $\tau$
satisfying (\ref{D1}) and if the bilinear form (\ref{bil}) makes
sense, then this gives the following bilinear identity for the KP
tau-functions:
\begin{equation}
\label{bil2}
Res_{z=0}\tau(t-[z^{-1}])e^{\sum (t_i-s_i)z^i}\tau(s+[z^{-1}])=0.
\end{equation}

\section{The Segal--Wilson Grassmannian}
An analytic approach to construct
wavefunctions of the KP-hierarchy is considered in \cite{SW}. Segal
and Wilson consider the Hilbert space
\[
H=\{ \sum_{n\in \mathbb{Z}} a_n z^n \mid a_n \in \mathbb{C} ,
\sum_{n\in\mathbb{Z}} \mid a_n\mid^2 < \infty\},
\]
with decomposition $H=H_+ \oplus H_-$, where
\[
H_+ = \{ \sum_{n\geq 0} a_n z^n \in H\} \quad \mbox{and}\quad H_- = \{
\sum_{n<0} a_n z^n \in H\}
\]
and inner product
$<\cdot \mid \cdot >$ given by
\[
< \sum_{n\in \mathbb{Z}} a_n z^n \mid \sum_{m\in \mathbb{Z}} b_m z^m >
  = \sum_{n\in \mathbb{Z}} a_n \overline{b_n} .
\]
To this decomposition is associated the Grassmannian $Gr(H)$ consisting
of all closed subspaces $W$ of $H$ such that the orthogonal projection
$p_+ : W\rightarrow H_+$ is Fredholm and the orthogonal projection
$p_- : W \rightarrow H_-$ is Hilbert-Schmidt. The connected components
of $Gr(H)$ are given by
\[
Gr^{(\ell)} (H) = \left\{
 W\in Gr(H) |\;
p_+:z^{-\ell}W\rightarrow  H_+ \;\;\mbox{has index zero}
\right\}.
\]
Each of these components
is a homogeneous space for the group $Gl_{res}^{(0)}(H)$ of all
bounded invertible operators $g:H\to H$ that decompose with respect to
$H=H_+ \oplus H_-$ as
\[
\left (
\begin{array}{cc}
a&b\\
c&d\\
\end{array}
\right ),
\]
where $a$ and $d$ are Fredholm operators and $b$ and $c$ are
Hilbert--Schmidt.
For each $N>1$, we consider the multiplicative group
\[
\Gamma_+(N) = \{ \exp (\sum_{i\geq 1} t_i z^i )\mid t_i \in \mathbb{C} ,\
\sum |t_i| N^i< \infty \}.
\]
equiped with the uniform norm. These groups are nested in a natural way
and  the inductive limit is denoted by $\Gamma_+$. The groups
$\Gamma_+(N)$ act by multiplication on $H$ and this gives a continuous
injection of $\Gamma_+(N)$ into $Gl_{res}^{(0)}(H)$. The commuting
flows from $\Gamma_+(N)$ lead to wavefunctions of the KP hierarchy for
which the product in (\ref{1.6h}) is real and not formal.

Now we take {}for $R$ the ring of meromorphic functions on $\Gamma_+(N)$
and {}for $\partial_n$ the partial derivative w.r.t. the parameter
$t_n$ of $\Gamma_+(N)$.
 For each $W\in Gr^{(\ell)} (H)$, let $\Gamma_+^W(N)$ be given by
\[
\Gamma_+^W(N)=\{\gamma(t)=\exp (\sum_{i=1}^\infty t_iz^i)\in \Gamma_+(N)|
\ p_+:\gamma^{-1}z^{-\ell}W\to  H_+\
\mbox{is a bijection}\}.
\]
In a similar way as in \cite{SW}, one shows that $\Gamma_+^W(N)$ is
nonempty. For $\gamma \in \Gamma_+^W(N)$, let $P_W(t,z)$ be $z^\ell$
times the inverse
image of $1$ under the projection $p_+:\gamma^{-1}z^{-\ell}W\to
H_+$. Then  we associate to $W$ an oscillating function
$\psi_W(t,z)=P_W(t,z)e^{\sum_{i=1}^\infty t_iz^i}$
of type $z^\ell$, which has the properties that it is
defined on a dense open subset of $\Gamma_+(N)$
and that its boundary value at $|z|=1$ belongs to $W$.
Moreover, it is known that the range of
$\psi_W$ spans a dense subspace of $W$.
A crucial property of the elements $\gamma(t)$ in this dense open subset of
$\Gamma_+(N)$ is that
\[
W\cap (z^{\ell} H_+)^{\perp} \gamma(t) = \{ 0 \} .
\]
By exploiting this property, one shows that $\psi_W$ is a wavefunction
for the KP hierarchy.
Hence, if we write $\psi_W(t,z)=
P_W(t,\partial) \cdot e^{\sum t_i z^i}$ with $P_W \in R[\partial ,
\partial^{-1})$, then $L_W = P_W \partial P_W^{-1}$ is a solution of
the KP-hierarchy. Each component of $Gr(H)$ generates in this way the
same set of solutions of the KP-hierarchy, so it would suffice, as is done
in \cite{SW}, to consider only $Gr^{(0)} (H)$. However, we need to
consider here all components.

Let $W\in Gr^{(\ell)} (H)$, then $W^\perp$ is a closed subspace of
$H$, moreover
$p_-:W^\perp\to H_-$ is a Fredholm operator of index $-\ell$ and
$p_+:W^\perp\to H_+$ is a Hilbert--Schmidt operator. Interchanging the
role of $H_+$ and $H_-$, we see that $W^\perp$ is a plane in the
adjoint Grassmannian $Gr^{*} (H)$ consisting of planes $U$ for which
$p_-$ is a Fredholm operator and $p_+$ is a Hilbert--Schmidt operator.
The connected components of $Gr^{*} (H)$ are also homogeneous spaces
of the group $Gl_{res}^{(0)}(H)$ introduced above.
On
$Gr^{*} (H)$ we consider however the commuting flows that are the
adjoint of the ones on $Gr (H)$. Namely, for each $N>1$, we take the group
\[
\Gamma_-(N) = \{ \exp (\sum_{i\geq 1} r_i z^{-i} )\mid r_i \in \mathbb{C} ,\
\sum |r_i| N^i< \infty \}.
\]
and as the analogue of $\Gamma_+^W(N)$ we take here
\[
\Gamma_-^{W^\perp}(N)=\{g(r)=\exp (\sum_{i=1}^\infty r_iz^{-i})\in \Gamma_-(N)|
\ p_-:g^{-1}z^{\ell}W^\perp \to  H_-\
\mbox{is a bijection}\}.
\]
One verifies directly that
\[
\Gamma_-^{W^\perp}(N)=\{ (\gamma^*)^{-1}|\gamma\in\Gamma_+^{W}(N)\}.
\]
In particular it is a non-empty open dense subset of $\Gamma_-(N)$.
Now we can repeat the construction of the wavefunction $\psi_W$ with
$W\in Gr(H)$ replaced by $W^\perp\in Gr^*(H)$ and $\Gamma_+(N)$ by
$\Gamma_-(N)$,
thus we obtain a function $\psi_{W^\perp}$
 of the form
\[
\begin{aligned}
\psi_{W^\perp}(r,z)=\psi_{W^\perp}(g(r),z)=&\{ \sum_{s \ge
\ell-1}\alpha_s(r)z^s\} g(r)\\
=&\hat \psi_{W^\perp}(g(r),z)g(r).
\end{aligned}
\]
Here $\alpha_{\ell-1}=1$ and $g(r)$ belongs to $\Gamma_-^{W^\perp}(N)$.
For all such $g(r)$ the boundary value of $\psi_{W^\perp}$ on $|z|=1$
belongs to $W^\perp$. This last property also holds for
\[
\begin{aligned}
\left (\frac {\partial}{\partial r_1}\right )^n
\psi_{W^\perp}(g(r),z)=&\{ \sum_{s \ge
\ell-1} \sum_{p=0}^n {n\choose p} \frac{\partial^{n-p}\alpha_s(r)}{\partial
r_1^{n-p}}
z^{s-p}\} g(r)\\
:=&\{ \sum_{s \ge
\ell-n-1}\alpha_s^{(n)}(r)z^s\} g(r).
\end{aligned}
\]
Since the boundary value on $|z|=1$ belongs to $W$, we get for all
$\gamma(t)\in\Gamma_+^{W}(N)$ and all $g(r)\in\Gamma_-^{W^\perp}(N)$
the equality
\begin{equation}
\label{2.1h}
<\psi_W(\gamma(t),z)|\left (\frac {\partial}{\partial r_1}\right )^n
\psi_{W^\perp}(g(r),z)>=0.
\end{equation}
Recall that the innerproduct is also equal to
\begin{equation}
\label{2.2h}
<f(z)|h(z)>=\frac 1{2\pi i}\int_{|z|=1}
f(z)\overline{h(z)}\frac{dz}{z}.
\end{equation}
For each
$g(r)\in\Gamma_-^{W^\perp}(N)$ there is a
$\gamma(s)\in\Gamma_+^{W}(N)$ such that $g(r)=\gamma(-\overline s)$.
Hence we get on $|z|=1$
\[
\begin{aligned}
\overline{\left (\frac {\partial}{\partial r_1}\right )^n
\psi_{W^\perp}(g(r),z)}&=\{ \sum_{p \ge
\ell-n-1}\overline{\alpha_p^{(n)}(r)}z^{-p}\} e^{\sum_{i=1}^\infty
\overline{r_i}z^i}\\
&=\{ \sum_{j \le
n-\ell+1}\overline{\alpha_{-j}^{(n)}(r)}z^j\} e^{\sum_{i=1}^\infty
\overline{r_i}z^i}\\
&=\{ \sum_{j \le
n-\ell+1}\overline{\alpha_{-j}^{(n)}(-\overline{s})}z^j\} e^{\sum_{i=1}^\infty
s_iz^i}\\
&=(-)^n\left (\frac {\partial}{\partial s_1}\right )^n\{ \sum_{j \le
-\ell+1}\overline{\alpha_{j}(-\overline{s})}z^j\gamma(s)\}
{}.
\end{aligned}
\]
In other words $\overline{\psi_{W^\perp}(\gamma(-\overline{s},z))}$ is
a dual oscillating function on $M^*$ of type $z^{1-\ell}$. If we
substitute this into formula (\ref{2.2h}) and we put $s_k=t_k$ for all
$k\ge 1$, then we obtain from (\ref{2.1h}) for all $n\ge 0$
\[
\mathfrak{R}(\psi_{W}(\gamma(t),z),\partial^n(z^{-1}
\overline{\psi_{W^\perp}(\gamma(-\overline{t},z)}))=0.
\]
So
$
z^{-1}
\overline{\psi_{W^\perp}(\gamma(-\overline{t},z))}$ satisfies the
requirements in Lemma \ref{l1h} and we can say
\begin{theorem}
\label{t2.1h}
The adjoint wavefunction of $\psi_{W}(\gamma(t),z)$ is
$
z^{-1}
\overline{\psi_{W^\perp}(\gamma(-\overline{t},z))}$.
\end{theorem}
A consequence of this theorem is that the adjoint wavefunction of a
wavefunction of the Segal--Wilson class, falls again in this class.
This is by no means obvious from the defining formula. Moreover, we
can give a meaning to the bilinear formula for the Segal--Wilson
wavefunctions , viz., for all relevant $t$ and $s$ one has
\begin{equation}
\label{2.3h}
<\psi_W(t,z)|\overline{z\psi_W^*(s,z)}>=0.
\end{equation}
We can use the fact that the boundary values of $\psi_W(t,z)$ on $|z|=1$
are lying dense in $W$ and that the boundary values of $\psi_W(s,z)$
at $|z|=1$
also belong to $H$ to characterize inclusions inside $Gr(H)$. This
gives you so-called modified equations. There holds namely
\begin{proposition}
\label{p1}
Let $V,\ W$ be planes in $Gr(H)$. Then $V\subset W$ (and hence
$W^\perp\subset V^\perp$) if and only if
\[
<\psi_V(t,z)|\overline{z\psi_W^*(s,z)}>=0.
\]
\end{proposition}
Next we recall how the tau-functions arise in the Segal-Wilson
setting. This starts with a convenient description of components
$Gr^{(\ell)}(H)$ of the Grassmannian. Let $\mathfrak{P}_\ell$ be the
collection of embeddings $w:z^\ell H_+\to H$ such that with respect to
the decomposition $H=(z^\ell H_+)\oplus(z^\ell H_+)^\perp$ the operator
$w$ has the form $w={{w_+}\choose{w_-}}$, with $w_-$ a Hilbert-Schmidt
operator and $w_+-\text{Id}$ a trace class operator. Then
$\mathfrak{P}_\ell$ is a natural way a fibre bundle over $Gr^{(\ell)}(H)$
with fiber the group
\[
\mathfrak{T}_\ell=\{ t\in\text{Aut}(z^\ell H_+)|t-
\text{Id of trace
class}\}.
\]
To lift the action of $GL^{(0)}_{res}(H)$ on $Gr^{(\ell)}(H)$ to one
on $\mathfrak{P}_\ell$, one has to pass to an extension $Gl$ of
$Gr^{(\ell)}(H)$. It is defined by
\[
Gl=\{(g,q)|g=\left (
\begin{array}{cc}
a&b\\
c&d\\
\end{array}
\right ),\ g\in GL^{(0)}_{res}(H), \ q\in \text{Aut}(z^\ell H_+),\
aq^{-1}-\text{Id is trace class}\}.
\]
This group acts by $w\mapsto gw
q^{-1}$ on $\mathfrak{P}_\ell$. The elements of both
$\Gamma_+(N)$
and $\Gamma_-(N)$ embed in a natural way into $Gl$ through
\[
\gamma_+=\left (
\begin{array}{cc}
a&b\\
0&d\\
\end{array}
\right )\mapsto (\gamma_+,a)\text{ and }
\gamma_-=\left (
\begin{array}{cc}
p&0\\
r&s\\
\end{array}
\right )\mapsto (\gamma_-,p).
\]
This we assume through this paper.
For each $w\in \mathfrak{P}_\ell$,
we define $\tau_w:Gl\to \mathbb{C}$
by
\[
\tau_w((g,q))=\det((g^{-1}wq)_+).
\]
If $t$ belongs to $\mathfrak{T}_\ell$, then there holds
$\tau_{w\circ t}=\det(t)\tau_w$. The restriction of $\tau_W$ to
$\Gamma_+(N)$ gives you the required tau-function. For, if
$q_\zeta=1-\frac{z}{\zeta}\in \Gamma_+(N)$, then there holds
\begin{theorem}
\label{tH1,2}
Let $W$ be a plane in $Gr^{(\ell)}(H)$ and let
$\psi_W(t,z)=P_W(t,z)e^{\sum t_iz^i}$ be the corresponding
wavefunction of type $z^\ell$. Then we have
\[
\begin{aligned}
(a)\quad&P
_W(t,\zeta)=\zeta^\ell\frac{\tau_w(\gamma(t)q_\zeta)}
{\tau_w(\gamma(t))}=\zeta^\ell\frac{\tau_W(t-[\zeta^{-1}])}
{\tau_W(t)}\quad\text{and}\\
(b)\quad &\text{the adjoint wavefunction }
\psi^*_W \text{ satisfies }\\
\ &\psi^*_W(t,\zeta)=\zeta^{-\ell}\frac{\tau_w(\gamma(t)q_\zeta^{-1})}
{\tau_w(\gamma(t))}e^{-\sum_{i=1}^\infty t_i\zeta^i}=
\zeta^{-\ell}\frac{\tau_W(t+[\zeta^{-1}])}
{\tau_W(t)}e^{-\sum_{i=1}^\infty t_i\zeta^i}
.\\
\end{aligned}
\]
\end{theorem}
Property (a) was shown in \cite{SW} and the relation in (b) is the
combination of theorem \ref{t1.1h} and the computations in \cite{HP}. This
theorem shows that in the Segal-Wilson setting we have the equations in
(\ref{D1}).

\section{Elementary B\"acklund--Darboux Transformations}

In this section we want to consider a certain type of
B\"acklund--Darboux transformations the so-called elementary ones. For
some literature on B\"acklund--Darboux transformations we refer the
reader to \cite{AvM1}, \cite{BHY}, \cite{MS},\cite{Oe1} and \cite{Oe2}.

Let $\psi$ be a wavefunction of type $z^\ell$, not necessarily in the
Segal--Wilson setting. Like in \cite{Oe1} and \cite{Oe2},
we define elementary B\"acklund--Darboux transformations
that produce new wavefunctions of type $z^{\ell\pm 1}$. For $q,r \in
R$, one
considers the first order differential
operators
$q\partial q^{-1}$ and $r^{-1}\partial r$. An obvious question now is:
When are  $q\partial q^{-1}\psi$ and $r^{-1}\partial^{-1} r \psi$
again wavefunctions? The answer is given by the following theorem. But
first notice that $r^{-1}\partial^{-1} r \psi$ is well and uniquely
defined.
If $\psi=P(z)e^{\sum t_iz^i}=P(\partial)e^{\sum t_iz^i}$, then
\[
\begin{array}{rcl}
r^{-1}\partial^{-1} r \psi&=&r^{-1}\partial^{-1} r P(\partial)e^{\sum
t_iz^i}\\
&=&r^{-1}\sum_{k=0}^\infty (-)^k\frac{\partial^k (r P(\partial))}{\partial x^k}
\partial^{-k-1}e^{\sum
t_iz^i}\\
&=&r^{-1}\sum_{k=0}^\infty (-)^k\frac{\partial^k (r P(z))}{\partial x^k}
z^{-k-1}e^{\sum
t_iz^i}.\\
\end{array}
\]
\begin{theorem}
  \label{t1}
Let $\psi$ be a wavefunction of type $z^\ell$ and let $\psi^*$ be its
adjoint wave function. Then the following holds:

\noindent
(a)\
Let  $q$ and $r\in R$ satisfy
\begin{eqnarray}
\label{2}
\partial_n (q)&=&(L^n)_+(q),\\
\label{3}
\partial_n (r)&=&-(L^n)_+^*(r)\quad\mbox{for all }n=1,2,\ldots
\end{eqnarray}
Then $q\partial  q^{-1}\psi$ (resp. $r^{-1}\partial^{-1}  r\psi$) is a
wavefunction of type $z^{\ell+1}$ (resp. $z^{\ell-1}$)
and
$-q^{-1}\partial^{-1}q\psi^*$ (resp. $-r\partial r^{-1}\psi^*$)
is its adjoint wavefunction.

\noindent
(b)\
If both $q_1$ and $q_2$ (resp. $r_1$ and $r_2$) satisfy (\ref{2})
(resp. (\ref{3})) such that
\[
q_1\partial  q_1^{-1}= q_2\partial   q_2^{-1}\quad (\mbox{resp. }
r_1^{-1}\partial r_1= r_2^{-1}\partial  r_2),
\]
then $q_1=\lambda q_2$ (resp. $r_1=\lambda r_2$)
for certain $\lambda\in\mathbb{C}$.
\end{theorem}
We say that $q$ (resp. $r$) is an {\it eigenfunction} (resp. {\it adjoint
eigenfunction}) of the Lax operator $L$ if it satisfies (\ref{2})
(resp. (\ref{3})).

Recall that there is a unique pseudodifferential operator $P$
such that $\psi = P(e^{\sum t_i z^i})$. It has the {}form
\begin{equation}
\label{4.0}
P = \partial^{\ell} + \sum_{j<\ell} p_j \partial^j = \{ 1 +
\sum_{s<0} p_{\ell + s} \partial^s \} \partial^{\ell},
\end{equation} which satisfies the Sato--Wilson equation (\ref{1.3.1}).
Let $\tau_q$ (resp. $\tau_r$) be the tau-function corresponding to
$q\partial q^{-1}\psi$ (resp. $r^{-1}\partial^{-1}r\psi$), by using formula
(\ref{D1}) one easily deduces that
\begin{equation}
\label{log}
\frac{\partial\log q}{\partial x}=\frac{\partial\log \tau_q/\tau}{\partial x}
\quad(\hbox{resp. }
\frac{\partial\log r}{\partial x}=\frac{\partial\log \tau_r/\tau}{\partial x}).
\end{equation}

Next we consider the operators $P_q$ and
$P_r$ defined by
\begin{equation}
  \label{1.3.2}
  P_q := q \partial q^{-1} P \quad \mbox{and}\quad P_r :=
  r^{-1} \partial^{-1} r P.
\end{equation}
To Prove part (a) of Theorem \ref{t1}, one has to show that the $P_q$ and the
$P_r$ also satisfy the
Sato-Wilson equations. To do so, we need some properties of the ring
$R[\partial , \partial^{-1})$ of pseudodifferential operators with
coefficients from $R$. We resume them in a lemma

\begin{lemma}
  \label{L1.3.1}
If $f$ belongs to $R$ and $Q$ to $R[\partial , \partial^{-1})$, then
the following identities hold
\begin{enumerate}
\item[(a)] $(Q f)_- = Q_- f$,
\item[(b)] $(fQ)_- = fQ_-$,
\item[(c)] Res$_{\partial}(Qf) =$ Res$_{\partial} (fQ) = f$ Res$_\partial (Q)$,
\item[(d)] $(\partial Q)_- = \partial Q_- -$ Res$_{\partial} (Q)$,
\item[(e)] $(Q\partial )_- = Q_- \partial-$ Res$_{\partial} (Q)$,
\item[(f)] $(Q\partial^{-1})_- = Q_- \partial^{-1} +$ Res$_{\partial}
  (Q\partial^{-1}) \partial^{-1}$,
\item[(g)] $(\partial^{-1} Q)_- = \partial^{-1} Q_- + \partial^{-1}$
  Res$_{\partial} (Q^* \partial^{-1})$,
\item[(h)]Res$_\partial(Qf\partial^{-1})=Q_+(f)$,
\item[(i)]Res$_\partial(\partial^{-1}fQ)=Q_+^*(f)$.
\end{enumerate}
\end{lemma}
Since the proof of this lemma consists of straight{}forward
calculations, we leave this to the reader. Now we can show

\begin{proposition}
  \label{P1.3.2}
The operators $P_q$ and $P_r$ satisfy the Sato-Wilson equations.
\end{proposition}

\noindent{\bf Proof.}
If we denote $ \frac{\partial}{\partial t_n}$ by $\partial_n$, then we
get {}for $P_q = q \partial q^{-1} P$ that
\[
\begin{array}[lcl]{lcl}
\partial_n (P_q)P_q^{-1} &=&
\partial_n (q \partial q^{-1}) q \partial^{-1} q^{-1} + q
\partial q^{-1} \partial_n (P) P^{-1} q \partial^{-1}
q^{-1}\\
&=& -q \partial q^{-1} (L^n)_- q \partial^{-1} q^{-1} +
\partial_n (q \partial q^{-1}) q \partial^{-1} q^{-1}.
\end{array}
\]
Now we apply successively the identities from Lemma \ref{L1.3.1} to the
first operator of the right-hand side
\[
\begin{array}[h]{lclc}
q \partial q^{-1} (L^n)_- q \partial^{-1} q^{-1} &=&
q \partial (q^{-1} L^n q)_- \partial^{-1} q^{-1} &=\\

q \partial (q^{-1} L^n q \partial^{-1})_- q^{-1} &-&
q \partial \mbox{Res}_{\partial} (q^{-1} L^n q
\partial^{-1})\partial^{-1} q^{-1} &=\\

q (\partial q^{-1} L^n q \partial^{-1})_- q^{-1} &+&
q \mbox{Res}_{\partial} (q^{-1} L^n q \partial^{-1}) q^{-1}
&-\\

q \partial \mbox{Res}_{\partial} (q^{-1} L^n q \partial^{-1})
\partial^{-1} q^{-1} &=&
(q \partial q^{-1} L^n q \partial^{-1} q^{-1})_- &+\\

q^{-1} \mbox{Res}_{\partial} (L^n q \partial^{-1}) &-&
q \partial q^{-1} \mbox{Res}_{\partial} (L^n q \partial^{-1})
\partial^{-1} q^{-1}\\

(q \partial q^{-1} L^n q \partial^{-1} q^{-1})_- &+& (L^n)_+
  (q) q^{-1} - q \partial q^{-1} (L^n)_+ (q) \partial^{-1} q^{-1}.
\end{array}
\]

On the other hand
\[
\partial_n (q \partial q^{-1}) q \partial^{-1} q^{-1} =
\partial_n (q) q^{-1} - q \partial q^{-2} \partial_n (q) q
\partial^{-1} q^{-1}.
\]
Thus we see that, if $\partial_n (q) = (L^n)_+ (q)$, the
operator $P_q$ satisfies the Sato-Wilson equation
\begin{equation}
  \label{1.3.3}
  \partial_n (P_q) P_q^{-1} = -(P_q \partial^n P_q^{-1})_- .
\end{equation}
{}For $P_r$, we proceed in a similar fashion
\[
\begin{array}[h]{l}
\partial_n (P_r) P_r^{-1} = -r^{-1} \partial^{-1} r (L^n)_- r
\partial r + \partial_n (r^{-1} \partial^{-1} r) r^{-1}
\partial r\\
= -r^{-1} \partial^{-1} (r L^n r^{-1})_- \partial r +
-\partial_n (r) r^{-1} + r^{-1} \partial^{-1} (\partial_n (r)
r^{-1})\partial r.
\end{array}
\]
Now we successively apply Lemma \ref{L1.3.1} (g) and (c) and (\ref{1.3.1}) to
the first term of the right hand side of this equation
\[
\begin{array}[h]{clclcl}
 -& r^{-1}\partial^{-1} (r L^n r^{-1})_- \partial r
&=& -r^{-1} \{(\partial^{-1} r L^n r^{-1} )_-
&-& \partial^{-1} \mbox{Res}_\partial (r^{-1} (L^n)^*_+ r \partial^{-1})\}
    \partial r\\
 =& -r^{-1} (\partial^{-1} r L^n r^{-1})_- \partial r
&+& r^{-1} \partial^{-1} r^{-1} (L^n)^*_+ (r) \partial r
& & \\
 =& -r^{-1} \{ (\partial^{-1} r L^n r^{-1} \partial)_-
&+& \mbox{Res}_{\partial} (\partial^{-1} r L^n r^{-1}) \} r
&+& r^{-1} \partial^{-1} r^{-1} (L^n)^*_+ (r) \partial r\\
 =& -(r^{-1} \partial^{-1} r L^n r^{-1}\partial r)_-
&-& r^{-1} (L^n)^*_+ (r)
&+& r^{-1} \partial^{-1} r^{-1} (L^n)^*_+ (r) \partial r.
\end{array}
\]
Since $\partial_n (r) = -(L^n)^* (r)$, we see that the last two
terms cancel $\partial_n (r^{-1} \partial r) r^{-1}
\partial r$ and thus we have obtained the Sato-Wilson equation {}for
$P_r$
\begin{equation}
  \label{1.3.5}
  \partial_n (P_r) P_r = -(P_r \partial^n P_r^{-1})_-.
\end{equation}
This concludes the proof of Proposition \ref{P1.3.2}.
\hfill{$\square$}
\\
\noindent
Thus  the proof of part (a) of Theorem \ref{t1} has been completed.
To prove part (b), we first calculate
\[
\partial\left (\frac{q_1}{q_2}\right )=\frac{q_1}{q_2}\frac{\partial(
q_1)}{q_1}
-\frac{q_1}{q_2}\frac{\partial( q_2)}{q_2}=0
\]
So $\partial^k(\frac{q_1}{q_2})=0$ and
$\partial^k(q_1)=\partial^k(\frac{q_1}{q_2}q_2)
=\frac{q_1}{q_2}\partial^k(q_2)$.
Now let $B_\ell=(L^\ell)_+$, then
\[
\begin{aligned}
\partial_n\left (\frac{q_1}{q_2}\right )
=&\frac{B_n(q_1)}{q_2}-\frac{q_1}{q_2^2}B_n(q_2)\\
=&\frac{1}{q_2}B_n(\frac{q_1}{q_2}q_2)-\frac{q_1}{q_2^2}B_n(q_2)\\
=&\frac{q_1}{q_2^2}B_n(q_2)-\frac{q_1}{q_2^2}B_n(q_2)\\
=&0
\end{aligned}
\]
for all $n=1,2,\ldots$. So we conclude that $q_1=\lambda q_2$ for a
certain $\lambda \in \mathbb{C}$.
Uniqueness for $r$ follows analogously.
This finishes the proof of Theorem \ref{t1}.

Suppose now
$q\partial q^{-1}\psi$
and  $r^{-1}\partial^{-1}r\psi$ are again wavefunctions, but $q$
is not an eigenfunction and $r$ is not an adjoint eigenfunction. Since $P_q$
and $P_r$ satisfy the Sato--Wilson equation,
\[
\begin{aligned}
\partial_n(q)-(L^n)_+(q)-q\partial q^{-1}\{
\partial_n(q)-(L^n)_+(q)\}\partial^{-1}&=0,\\
\partial_n(r)+(L^n)_+^*(r)-\partial\{\partial_n(r)+(L^n)_+^*(r)\}r^{-1}\partial
r&=0
\end{aligned}
\]
and thus
\[
\begin{aligned}
\partial(q^{-1}\{
\partial_n(q)-(L^n)_+(q)\})&=0,\\
\partial(r^{-1}\{\partial_n(r)+(L^n)_+^*(r)\} )&=0.
\end{aligned}
\]
So
\begin{equation}
\begin{aligned}
\partial_n(q)-(L^n)_+(q)&=f_n(t_2,t_3,\ldots)q,\\
\partial_n(r)+(L^n)_+^*(r)&=g_n(t_2,t_3,\ldots)r.
\end{aligned}
\end{equation}
Let again $B_\ell=(L^\ell)_+$, and write down the compatibility conditions
for $q$, i.e.,
$\partial_n(\partial_\ell(q)=\partial_\ell(\partial_n(q))$. One has
\[
\begin{aligned}
\partial_n(\partial_\ell(q))=&\partial_n(B_\ell(q)+f_\ell q)\\
=&\partial_n(B_\ell)(q)+B_\ell(\partial_n(q))+\partial_n(f_\ell)q
+f_\ell\partial_n(q)\\
=&\partial_n(B_\ell)(q)+B_\ell B_n(q)+B_\ell(f_nq)+
\partial_n(f_\ell)q+f_\ell B_n(q)+f_\ell f_n q
\end{aligned}
\]
and a similar relation with $n$ and $\ell$ interchanged, so we find
\[
\begin{aligned}
(\partial_n(B_\ell)-\partial_\ell(B_n)+B_\ell B_n-B_nB_\ell)(q)+&\\
 B_\ell(f_nq)-B_n(f_\ell
q)+
f_\ell B_n(q)-f_nB_\ell(q)+\\
\partial_n(f_\ell)q-\partial_\ell(f_n)q&=0.
\end{aligned}
\]
Because the Zakharov--Shabat  equations (\ref{1.1.2a}) hold for
the $B_\ell$'s
and $f_m$ is
independent of $x=t_1$ one finds that
\[
\partial_n(f_\ell)=\partial_\ell(f_n)
\]
and hence if it is possible to integrate
to all times $t_j$ one can find a function $F(t)$ such that
\begin{equation}
\label{f1}
f_n(t)=F(t)^{-1}\partial_nF(t).
\end{equation}
Now choose
\[
\hat q=F(t)^{-1}q,
\] then $\partial_n(\hat q)=B_n(\hat q)$. A similar argument holds for
$r$.
So we have under certain conditions, viz. certain integrability of functions,
shown that one can find an eigenfunction $\hat q$ (resp.
adjoint eigenfunction $\hat r$) which produces
the same elementary
B\"acklund--Darboux transformation, i.e.
\[
q\partial q^{-1}=\hat q\partial \hat q^{-1}\quad\hbox{resp. }
r\partial r^{-1}=\hat r\partial \hat r^{-1}.
\]
We will show that in the Segal--Wilson case the above conditions are satisfied.
In that case we will explicitly construct the
(adjoint) eigenfunctions $\hat q$ and $\hat r$.

Clearly the B\"acklund--Darboux operators of Theorem \ref{t1} are
invertible. Moreover, we prove the following useful
lemma.
\begin{lemma}
\label{l1}
Let $q,\ r\in R$, then $q$ (resp. $r$) satisfies (\ref{2}) (resp.
(\ref{3}))
if and only if $q^{-1}$ (resp. $r^{-1}$)
satisfies
\[
\begin{array}{rcl}
\partial_n (  q^{-1})
& = &-(q\partial  q^{-1}L^nq\partial^{-1}q^{-1})_+^* (q^{-1}),\\
\partial_n (  r^{-1})
& = &\phantom{-}(r^{-1}\partial^{-1}  rL^nr^{-1}\partial r)_+(r^{-1}),\quad
\mbox{respectively}.
\end{array}
\]
\end{lemma}

\noindent{\bf Proof.}
Suppose (\ref{2}) holds, then
\[
\begin{aligned}
\partial_n(q^{-1})=&-q^{-2}(L^n)_+(q)\\
=&-q^{-2}Res_\partial(\partial^{-1}L^n)^*\\
=&-Res_\partial(q^{-1}\partial^{-1}qL^{n*}q^{-1})\\
=&-Res_\partial(q^{-1}\partial^{-1}qL^{n*}q^{-1}\partial
qq^{-1}\partial^{-1})\\
=&-(q^{-1}\partial^{-1}qL^{n*} q^{-1}\partial q)_+(q^{-1})\\
=&-(q\partial  q^{-1}L^nq\partial^{-1}q^{-1})_+^* (q^{-1}).
\end{aligned}
\]
The second equation of this Lemma is proven in a similar fashion.
\hfill$\square$
\section{Trivial B\"acklund--Darboux Transformations}
Besides the elementary B\"acklund--Darboux transformations
one also has trivial B\"acklund--Darboux transformations.
These transformations do not change the Lax operator $L$.
They consist of multiplying a wavefunction with a function of the form
\begin{equation}
\label{-3.1}
R(z)=z^k(1+\sum_{j=1}^\infty r_j z^{-j})\quad\hbox{with }
r_j\in\mathbb{C}.
\end{equation}
If $\psi(t,z)=P(t,\partial)e^{\sum_it_iz^i}$ is a wavefunction
of type $z^\ell$, then
\[
\begin{aligned}
\psi_R(t,z)=&R(z)\psi(t,z)=P(t,\partial)R(\partial)e^{\sum_it_iz^i}\\
=&z^k(1+\sum_{j=1}^\infty r_jz^{-j})\psi(t,z)\\
\end{aligned}
\]
is the new wavefunction of type $z^{k+\ell}$. Its adjoint wavefunction is
of the form
\[
\psi_R^*(t,z)=P(t,\partial)^{*-1}R(\partial)^{*-1}e^{-\sum_it_iz^i}
=R(z)^{-1}\psi^*(t,z).
\]
Since all $r_j$'s are constants, this
does not change the form of $L$, but it
clearly changes the tau-function.
Rewrite $R(z)$ as in (\ref{-3.1}) as follows
\begin{equation}
\label{-3.2}
R(z)=z^ke^{-\sum_{j=1}^\infty \frac{q_j}{j}z^{-j}},
\end{equation}
then
\[
\begin{aligned}
\psi_R(t,z)=&z^k e^{-\sum_{j=1}^\infty \frac{q_j}{j}z^{-j}}
\psi(t,z)\\
=&z^{k+\ell}\frac{\tau(t-[z^{-1}])}{\tau(t)}
e^{-\sum_{j=1}^\infty \frac{q_j}{j}z^{-j}}
e^{\sum_{i=1}^\infty t_i z^i}\\
=&z^{k+\ell}\frac{\tau(t-[z^{-1}])
e^{\sum_{j=1}^\infty q_j(t_j-\frac{z^{-j}}{j})}}
{\tau(t)e^{\sum_{j=1}^\infty q_jt_j}}
e^{\sum_{i=1}^\infty t_i z^i},
\end{aligned}
\]
which suggests that $\tau_R(t):=\tau(t)e^{\sum_{j=1}^\infty q_jt_j}$
is the to $\psi_R$ corresponding tau-function.
We will show this in the Segal-Wilson context. First of all, one
notices that the multiplication by $z^k$ does not change the
tau-function. Hence we may assume that $k=0$ and we consider the
action of
\[
\rho=\exp ({-\sum_{j=1}^\infty \frac{q_j}{j}z^{-j}})\in \Gamma_-(N)
\]
on $\mathfrak{P}_\ell$.
The operator $\rho$ decomposes with respect to
$H=z^\ell H_+\oplus(z^\ell H_+)^\perp$
as
$\rho=\left (
\begin{array}{cc}
\alpha&0\\
\beta&\delta\\
\end{array}
\right )$.
Hence its action on
$w=\left (\begin{array}{c}w_+\\ w_-\\\end{array}\right )\in\mathfrak{P}_\ell$
is
\[
\rho\circ w\circ \alpha^{-1}=\left (
\begin{array}{c}
\alpha w_+\alpha^{-1}\\
(\beta w_+ + \delta w_-)\alpha^{-1}\\
\end{array}
\right ).
\]
Let $\gamma^{-1}=\exp(-\sum_i t_iz^i)\in \Gamma_+(N)$ decompose
with respect to
$H=z^\ell H_+\oplus(z^\ell H_+)^\perp$
as $\gamma^{-1}=\left (
\begin{array}{cc}
a&b\\
0&d\\
\end{array}
\right )$.
Then the tau-function corresponding to $\rho W$, where $W=\text{\rm Im}(w)$,
$w
\in\mathfrak{P}_\ell$, is by definition
\[
\tau_{\rho\circ w\circ \alpha^{-1}}(\gamma)=
\det((\gamma^{-1} \rho w\alpha^{-1} a^{-1})_+).
\]
Since $\gamma^{-1}$ and $\rho$ commute we have the relation
\[
\alpha a=a\alpha +b\beta\quad\text{or equivalently }
\alpha a\alpha^{-1} a^{-1}=I+b\beta\alpha^{-1} a^{-1},
\]
so that we see that the operator
$\alpha a\alpha^{-1} a^{-1}$ is of the form ``{\it identity +
trace-class}'' and hence has a determinant. Thus we get for the
tau-function of $\rho W$ that
\[
\begin{aligned}
\tau_{\rho\circ w\circ \alpha^{-1}}(\gamma)=&
\det((\gamma^{-1} \rho w a^{-1} \alpha^{-1})_+)
\det (\alpha a\alpha^{-1} a^{-1})\\
=&\det((\rho \gamma^{-1} w a^{-1} \alpha^{-1})_+)
\det (\alpha a\alpha^{-1} a^{-1})\\
=&\tau_w(\gamma)\det (\alpha a\alpha^{-1} a^{-1})\\
\end{aligned}.
\]
Hence we merely have to show that
\[
B(\alpha,a)=\det (\alpha a\alpha^{-1} a^{-1})
\]
is the required function. This we will prove in a few steps. First one
notices
that if
$\rho_1=\left (
\begin{array}{cc}
\alpha_1&0\\
\beta_1&\delta_1\\
\end{array}
\right )$ and
$\rho_2=\left (
\begin{array}{cc}
\alpha_2&0\\
\beta_2&\delta_2\\
\end{array}
\right )$ are in $\Gamma_-(N)$, then
\[
\begin{aligned}
B(\alpha_1\alpha_2, a)=&\det (\alpha_1\alpha_2
a\alpha_2^{-1}\alpha_1^{-1}a^{-1})
=\det (\alpha_2 a\alpha_2^{-1}a^{-1} a\alpha_1^{-1}a^{-1}\alpha_1)\\
=&\det (\alpha_2
a\alpha_2^{-1}a^{-1})\det(a\alpha_1^{-1}a^{-1}\alpha_1)\\
=&B(\alpha_2,a)B(\alpha_1,a).\\
\end{aligned}
\]
Likewise one shows for $\gamma_i^{-1}=\left (
\begin{array}{cc}
a_i&b_i\\
0&d_i\\
\end{array}
\right )\in \Gamma_+(N)$
that
\[
B(\alpha, a_1a_2)=B(\alpha,a_1)B(\alpha,a_2).
\]
Hence if we put
$\rho_i=\exp(-\frac{q_i}{i}z^{-i})=\left (
\begin{array}{cc}
\alpha_i&0\\
\beta_i&\delta_i\\
\end{array}
\right )$,
then $\rho=\lim_{N\to\infty}\rho_1\rho_2\cdots\rho_N$.
The form $B$ is continuous in $\alpha$, so that we get
\[
B(\alpha,a)=\prod_{i=1}^\infty B(\alpha_i,a)
\]
and thus we merely have to prove the formula for $\rho_i$. If
\[
\gamma_i^{-1}=
\exp(-t_iz^i)
=\left (
\begin{array}{cc}
a_i&b_i\\
0&d_i\\
\end{array}
\right )\in \Gamma_+(N),
\]
then $\gamma=\lim_{N\to\infty}\gamma_1\gamma_2\cdots\gamma_N$. As $B$
is also  continuous and multiplicative in $a$, we end up with the
formula
\[
B(\alpha,a)=\prod_{i,j\ge1}B(\alpha_i,a_j).
\]
Therefore we only have to prove the formula for $\rho=\rho_i$ and
$\gamma=\gamma_j$. Clearly the map $q_i\to B(\alpha_i,a_j)$
defines a continuous morphism $\chi_i^{(j)}$ from $\mathbb{C}$ to
$\mathbb{C}^\times$ and therefore it has the form
\[
\chi_i^{(j)}(q_i)=e^{\beta_{ij}q_i},
\]
for some $\beta_{ij}\in \mathbb{C}$, depending on $\gamma_j$. Likewise
the map $t_j\to B(\alpha_i,a_j)$ defines a continuous morphism
$\phi_j^{(i)}$
from $\mathbb{C}$ to
$\mathbb{C}^\times$ that can be written as
\[
\phi_j^{(i)}(t_j)=e^{\alpha_{ij}t_j},
\]
for some $\alpha_{ij}\in \mathbb{C}$, depending on $\rho_j$.
Hence we can say that for all $i\ge 1$ and $j\ge 1$
\[
B(\alpha_i, a_j)=e^{a_{ij}t_iq_j}\quad\text{ with }a_{ij}\in \mathbb{C}.
\]
The $\{a_{ij}\}$ can easily be determined by computing the lowest
non-trivial term in $B(\alpha_i, a_j)$. Since
\[
b_j=\left (
\begin{array}{cccccc}
\vdots& & & & & \\
0& & & &0& \\
-t_j&\ddots& & & & \\
0&\ddots&\ddots& & & \\
\vdots &\ddots&\ddots&\ddots& & \\
0&\cdots&0&-t_j&0&\cdots\\
\end{array}
\right )
\quad\text{and }
\beta_j=
\left (
\begin{array}{cccccc}
\cdots&0&-\frac{q_i}{i}&0&\cdots&0\\
&&\ddots&\ddots&\ddots&\vdots\\
&&&\ddots&\ddots&0\\
&&&&\ddots&-\frac{q_i}{i}\\
&0&&&&0\\
&&&&&\vdots\\
\end{array}
\right ),
\]
\[
\begin{aligned}
B(\alpha_i,a_j)=&
\det(I+\alpha_i^{-1}a_j^{-1}b_j\beta_i)\\
=&1+\sum_{n\ge 1}\text{\rm Tr}
(\Lambda^n(\alpha_i^{-1}a_j^{-1}b_j\beta_i))\\
\end{aligned}
\]
we see that only $\text{\rm Tr}
(\alpha_i^{-1}a_j^{-1}b_j\beta_i)$ contributes to the lowest
non-trivial term. For the same reason the lowest non-trivial term of
$\text{\rm Tr}
(\alpha_i^{-1}a_j^{-1}b_j\beta_i)$ is equal to that of
$\text{\rm Tr}
(b_j\beta_i)$.
A straightforward calculation gives the required result, viz.,
$a_{ij}=0$ if $i\ne j$ and $a_{ii}=1$.
We resume the results of this computation in a theorem.
\begin{theorem}
\label{t4.1h}
Let $\rho$ be the element $\exp(\sum_{j\ge 1}-\frac{q_j}{j}z^{-j})$ in
$\Gamma_-(N)$. Then the tau-function corresponding to the plane $\rho
W$ is given by
\[
\tau_{\rho W}(t)=
\tau_{\rho\circ w\circ \alpha^{-1}}(\gamma)=
\exp(\sum_{j\ge 1}\frac{q_j}{j}t_j)\tau_w(\gamma)
=\exp(\sum_{j\ge 1}\frac{q_j}{j}t_j)\tau_W(t).
\]
\end{theorem}

\section{Geometrical Interpretation of
B\"acklund--Darboux Transformations}

In this section we want to interpret
the trivial and elementary B\"acklund--Darboux transformations
on the Segal--Wilson Grassmannian. As in section 2 we consider the
flows in $\Gamma_+(N)$. First consider the trivial transformations.
Let $\psi_W$ be a wavefunction of type $z^\ell$,
corresponding to $W\in Gr^{(\ell)}(H)$.
Then clearly multiplication by $
R(z)=z^ke^{-\sum_{j=1}^\infty \frac{q_j}{j}z^{-j}}$
maps $\psi_W(t,z)$ into $\psi_R(t,z)=R(z)\psi_W(t,z)$.
Since this has to be a Segal--Wilson wavefunction of type $z^{k+\ell}$,
we have to assume that $\sum_{j=1}^\infty |q_j|\frac{N^j}{j}<\infty$. It is
then straightforward to see that $\psi_R$ is the wavefunction
corresponding to $R(z)W\in Gr^{(k+\ell)}(H)$, hence the trivial
B\"acklund--Darboux
transformation
maps W into $R(z)W$.

Next consider the elementary B\"acklund--Darboux transformations.
The following consequence of
Theorem \ref{t1} is more
or less well known (see e.g. \cite{HP} and \cite{AvM1}).

\begin{corollary}
  \label{C4.3}
The functions $\psi_q=q\partial q^{-1}\psi_W$ and
$\psi_r=r^{-1}\partial^{-1}r\psi_W$ (resp.
$\psi^*_q=-q^{-1}\partial^{-1} q\psi_W^*$ and $\psi_r^*=-r\partial
r^{-1}\psi_W^*$) are (adjoint) wavefunctions of
planes $W_q$ and $W_r$. Moreover we have the following
codimension 1 inclusions:
\[
W_q \subset W,\quad  W^\perp\subset W_q^\perp,\quad W\subset W_r,
\quad \mbox{and}\quad W_r^\perp\subset W^\perp
\]
\end{corollary}

\noindent {\bf Proof.}
{}From the Sato-Wilson equations one deduces directly that {}for all
$n\geq 1$,
\[
\partial_n \psi_q = (P_q \partial^n P_q^{-1})_+ \psi_q
\quad \mbox{and}\quad
\partial_n \psi_r = (P_r \partial^n P_r^{-1})_+ \psi_r.
\]
This shows the first part of the claim.
Consider the following subspace in $Gr(H)$:
\[
W_q = \mbox{the closure of Span} \{ \psi_q (t,z)\}.
\]
The inclusions between the
spaces $W$ and $W_q$ follows from the first relation of (\ref{1.3.2})
and the fact that the values of a wavefunction corresponding to an
element of $Gr(H)$ are lying dense in that space. Since {}for a suitable
$\gamma$ in $\Gamma_+(N)$ the orthogonal projections of $\gamma^{-1}
W_q$ on $z^{\ell} H_+$ resp. $\gamma^{-1} W$ on $z^{\ell + 1} H_+$
have a one dimensional kernel, one obtains the codimension one
result.
For the inclusions between the
spaces $W$ and $W_r$ we consider the adjoint wavefunctions
$\psi^*_W=P_W^{*-1}e^{-\sum t_i z^i}$ and $\psi_r^*=-r\partial r^{-1}\psi^*_W$.
Since the complex conjugate $\overline{z\psi_{W}^* (t,z)}$ of $z\psi^*_W(t,z)$
corresponds to the space $W^\perp$, the same argument
as before shows the codimension $1$ inclusion:
\[
W_r^\perp:=\mbox{the closure of Span}\{\overline{z\psi_r^* (t,z)}\}\subset
W^\perp.
\]
Hence $\psi_r (t,z)$ corresponds to
$W_r
= {\mbox{the closure of Span}\{\psi_r (t,z)\}}$.
This concludes the proof of the corollary.
\hfill{$\square$}

Now suppose we are in the situation of the above Corollary. Let
$s(z)\in W\cap W_q^\perp$, then  $W_q=\{ w\in W|<w|s(z)>=0\}$ and
\[
0=<\psi_q(t,z)|s(z)>=<\frac{\partial\psi_W(t,z)}{\partial x}-q(t)^{-1}
\frac{\partial q(t)}{\partial x}\psi_W(t,z)|s(z)>.
\]
Hence,
\[
\frac{\partial \log q}{\partial x}
=\frac{\partial \log <\psi_W(t,z)|s(z)>}{\partial x}.
\]
Since both $q(t)$ and $<\psi_W(t,z)|s(z)>$ are eigenfunctions of $L$ that
define the same B\"acklund--Darboux transformation, one deduces from
Theorem \ref{t1} that
\[
q(t)=\lambda <\psi_W(t,z)|s(z)>
\quad \mbox{for certain }
\lambda\in\mathbb{C}^\times.
\]
Now notice two things. First
that we can also replace $s(z)\in W\cap W_q$ by
$\tilde s(z)=s(z)+t(z)$ with $t(z)\in W^\perp$.
And next that if $\partial +u$ is a B\"acklund-Darboux transformation
which maps $W$ into a subspace $U$, then for $s(z)\in W\cap U^\perp$,
$q(t)=<\psi_W(t,z)|s(z)>$ gives the eigenfunction such that
$\partial+u=q\partial q^{-1}$.

For the adjoint  B\"acklund--Darboux transformation, one takes
$t(z)\in W^\perp\cap W_r$, then
$W_r^\perp=\{w\in W^\perp|<t(z)|w>=0\}$,
$<z\psi_W^*(t,z)|\overline{t(z)}>=0$ and in a similar way as for $q$ one obtains
that
\[
r(t)=\lambda <z\psi_W^*(t,z)|\overline{t(z)}>=\lambda
<t(z)|\overline{z\psi_W^*(t,z)}>
\quad \mbox{for certain }
\lambda\in\mathbb{C}^\times.
\]
Again it is obvious that we can replace $t(z)$ by
$\tilde t(z)=s(z)+t(z)$ with $s(z)\in W$  and that this construction
gives the adjoint eigenfunction.

So we conclude that one has a geometric interpretation for the
elementary B\"acklund--Darboux transformation. It has its origin on
the Segal--Wilson Grassmannian, viz.,
\begin{proposition}
\label{p2}
Let $W_q$ and $W_r$ be  the planes of Corollary \ref{C4.3}, then
\[
q(t)=<\psi_W(t,z)|s(z)>\quad\mbox{and}\quad
r(t)=<t(z)|\overline{z\psi_W^*(t,z)}>
\] for certain $s(z)\in W\cap W_q^\perp$
and $t(z)\in W^\perp\cap W_r$.
\end{proposition}
The next step in analyzing the situation in $Gr(H)$ from proposition
\ref{p2},
\[
W_q\subset W\subset W_r,\ \dim (W_r/W)=\dim(W/W_q)=1,
\]
is to determine the tau-function corresponding to $W_q$ resp. $W_r$
and
their relation to that of $W$.
Let $q(t)\partial q(t)^{-1}$ (resp.
$-r(t)\partial r(t)^{-1}$) act on the wavefunction $\psi_W(t,z)$
(resp. adjoint wavefunction $\psi_W^*(t,z)$ and compare this with
$\psi_q(t,z)$ (resp. $\psi_r^*(t,z)$).
We thus obtain
\begin{equation}
\begin{aligned}
z P_{W_q}(t,z)&=\frac{\partial P_W(t,z)}{\partial x}+(z
-q^{-1}\frac{\partial q}{\partial x})P_W(t,z)\\
zP_{W_r}^{*-1}(t,-z)&=-\frac{\partial
P_W^{*-1}(t,-z)}{\partial x}
+(z
+r^{-1}\frac{\partial r}{\partial x})P_W^{*-1}(t,-z)
\end{aligned}
\end{equation}
Comparing the coefficients of $z$ we find that
\begin{equation}
\frac{\partial \log q}{\partial x}=\frac{\partial\log
\tau_{W_q}/\tau_W }{\partial x}
\quad\mbox{and}\quad
\frac{\partial \log r}{\partial x}=\frac{\partial\log
\tau_{W_r}/\tau_W }{\partial x}.
\end{equation}
This suggests the following relations
\[
\tau_{W_q}(t)=\lambda q(t)\tau_W(t)\quad\text{and }
\tau_{W_r}(t)=\mu r(t)\tau_W(t),
\]
with $\lambda,\ \mu\in \mathbb{C}^\times$.
Since the tau-function of a plane $W$ in $Gr(H)$ is only determined up
to a constant, one may assume that the constants $\lambda$ and $\mu$ are
equal to 1.

We will carry out the computations for $W_r$. Then that for $W_q$ is
an easy consequence of it. As above let $t(z)\in H$ be such that
\[
r(t)=<z\psi_W^*|\overline{t(z)}>=<t(z)|
\psi_{W^\perp}(\gamma(-\overline{t}),z)>.
\]
If $w:z^\ell H_+\to W$ is in $\mathfrak{P}_\ell$, then we can extend it to
an embedding $w_r:z^{\ell-1}H_+\to W_r$ by putting
$w_r(z^{\ell-1})=t(z)$ and
$w_r(f)=w(f)$ for $f\in z^\ell H_+$. Then $w_r$ belongs to
$\mathfrak{P}_{\ell-1}$. Next we decompose $w_r$ with respect to
$H=z^\ell H_+\oplus \mathbb{C}z^{\ell-1}\oplus (z^{\ell-1}H_+)^\perp$
and $w$ with respect to $H=z^\ell H_+\oplus  (z^{\ell}H_+)^\perp$,
then we get
\[
w_r=\left (
\begin{array}{cc}
w_{++}&w_{+0}\\
w_{0+}&w_{00}\\
w_{-+}&w_{-0}\\
\end{array}
\right )\quad\text{and }
w=\left (
\begin{array}{c}
w_+\\
w_-\\
\end{array}
\right )\quad\text{with }w_+=w_{++}\quad\text{and }w_-=\left (
\begin{array}{c}
w_{0+}\\
w_{-+}\\
\end{array}
\right ).
\]
We do the same with the operator $\gamma^{-1}$ from $\Gamma_+(N)$.
This gives you respectively
\[
\gamma^{-1}=\left (
\begin{array}{ccc}
a_{++}&a_{+0}&a_{+-}\\
0&1&a_{0-}\\
0&0&a_{--}\\
\end{array}\right )
=
\left (
\begin{array}{cc}
a_{++}&b\\
0&c\\
\end{array}
\right )\quad\text{with } b=(a_{+0}\ a_{+-})\quad\text{and }
c=\left (
\begin{array}{cc}
1&a_{0-}\\
0&a_{--}\\
\end{array}\right ).
\]
Before computing $\tau_{w_r}$, we notice that
\[
\left (
\begin{array}{cc}
a_{++}&0\\
0&1\\
\end{array}
\right )
\left (
\begin{array}{cc}
a_{++}^{-1}&-a_{++}^{-1}a_{+0}\\
0&1\\
\end{array}
\right )-I=\left (
\begin{array}{cc}
0&-a_{+0}\\
0&0\\
\end{array}
\right )
\]
is a 1-dimensional operator of zero-trace. Hence we have
\[
\det \left(
\left (
\begin{array}{cc}
a_{++}&0\\
0&1\\
\end{array}
\right )
\left (
\begin{array}{cc}
a_{++}^{-1}&-a_{++}^{-1}a_{+0}\\
0&1\\
\end{array}
\right ) \right )=1,
\]
so that it suffices to compute the projection of
$\gamma^{-1} w_r \left (
\begin{array}{cc}
a_{++}&0\\
0&1\\
\end{array}
\right )^{-1}$ onto $z^{\ell-1}H_+$:
\[
\left (
\begin{array}{cc}
a_{++}w_+a_{++}^{-1}+bw_-a_{++}^{-1}&a_{++}w_{+0}+a_{+0}w_{00}+a_{+-}w_{-0}\\
(w_{0+}+a_{0-}w_{-+})a_{++}^{-1}&w_{00}+a_{0-}w_{-0}\\
\end{array}
\right )=
\left (
\begin{array}{cc}
A&\beta\\
\alpha&\delta\\
\end{array}
\right ).
\]
As $A$ is equal to $(\gamma^{-1}wa_{++}^{-1})_+$ and $\gamma$ is
chosen such that this operator is invertible, we can decompose
$
\left (
\begin{array}{cc}
A&\beta\\
\alpha&\delta\\
\end{array}
\right )
$
as follows
\[
\left (
\begin{array}{cc}
A&\beta\\
\alpha&\delta\\
\end{array}
\right )=\left (
\begin{array}{cc}
1&0\\
\alpha A^{-1}&1\\
\end{array}
\right )\left (
\begin{array}{cc}
A&0\\
0&\delta-\alpha A^{-1}\beta\\
\end{array}
\right )\left (
\begin{array}{cc}
1&A^{-1}\beta\\
0&1\\
\end{array}
\right ).
\]
{}From this we see that
\[
\tau_{w_r}(\gamma)=\det\left (\left (
\begin{array}{cc}
A&\beta\\
\alpha&\delta\\
\end{array}
\right )\right )= \det (A)(\delta-\alpha A^{-1}\beta)
=\tau_w(\gamma)(\delta-\alpha A^{-1}\beta).
\]
Hence we merely have to show that $r(t)=\delta-\alpha A^{-1}\beta$. We
know that $g(r)=\gamma(-t)^*$ with
$\gamma(t)\in\Gamma_+^W(N)$ and thus we get
\[
r(t)=<\gamma(t)t(z)|\hat\psi_{W^\perp}(\gamma(-t)^*,z)>.
\]
By definition one has
\[
\hat\psi_{W^\perp}(\gamma(-t)^*,z)=z^{\ell-1}+\sum_{k\ge \ell}\psi_k
(\gamma(-t)^*)z^k.
\]
Therefore, if $\beta=\sum_{k\ge \ell}\beta_k(t)z^k$, then the
expression for $r(t)$ becomes
\[
\begin{aligned}
r(t)=&<\delta z^{\ell-1}+\sum_{k\ge \ell}\beta_k(t)z^k|
z^{\ell-1}+\sum_{j\ge \ell}\psi_j (\gamma(-t)^*)z^j>\\
=&\delta+\sum_{k\ge \ell}\beta_k(t)
\overline{\psi_k (\gamma(-t)^*)}.\\
\end{aligned}
\]
{}From the computation of $\hat\psi_{W^\perp}$ in \cite{HP}, we see
that, if $\alpha A^{-1}(z^{k-1})=\alpha_kz^{\ell-1}$ for all
$k\ge\ell$, then $\psi_k (\gamma(-t)^*)=-\overline{\alpha_k}$.
In other words, there holds
\[
r(t)=\delta-\alpha A^{-1}\beta.
\]
Next we treat the case $W_q\subset W$ and $\dim (W/W_q)=1$. Let
$s(z)\in W$ be such that $W_q=\{ w\in W|<w|s(z)>=0\}$ and
$q(t)=<\psi_W(t,z)|s(z)>$.
Let $w_0\in\mathfrak{P}_{\ell+1}$ be such that
$w_0(z^{\ell+1}H_+)=W_q$,
then $w=(w_0\  s(z))$ belongs to $\mathfrak{P}_{\ell}$ and $w(z^\ell
H_+)=W$.
According to the foregoing result, we have
\[
\tau_w(t)=r(t)\tau_{w_0}(t)\quad\text{with
}r(t)=<s(z)|\psi_{W_q^\perp}(\gamma(-\overline{t}))>.
\]
In particular, one has the relations
\[
P_W=r^{-1}\partial^{-1}rP_{W_q}\quad\text{and } P_{W_q}=q\partial
q^{-1}P_W,
\]
so that we may conclude
\[
r^{-1}\partial r=q\partial q^{-1}.
\]
According to Lemma \ref{l1} $r^{-1}$ satisfies
\[
\partial_n(r^{-1})=(r^{-1}\partial^{-1}
rP_{W_q}\partial^nP_{W_q}^{-1}r^{-1}\partial
r)_+(r^{-1})= (L_W^n)_+(r^{-1})
\]
like $q$. Hence by Theorem \ref{t1} we may conclude that $q(t)=\lambda
r(t)^{-1}$
with $\lambda\in \mathbb{C}^\times$.
We resume the forgoing result in the form of a theorem.
\begin{theorem}
\label{t5.1h}
Let $W,W_q,W_r\in Gr(H)$ be such that $W_q\subset
W\subset W_r$ and $\dim(W_r/W)=\dim(W/W_q)=1$. Then the tau-function
of $ W_r$
is equal to $\tau_{W_r}(t)=r(t)\tau_W(t)$, with
$r(t)=<v|\psi_{W^\perp}(\overline{t},z)>$
for some $v\in W_r\backslash W$ and the tau-function of $W_q$ is equal to
$\tau_{W_q}(t)=q(t)\tau_w(t)$, where $q(t)=<\psi_W(t,z)|s>$ fore some $s\in
W\cap (W_q)^\perp$.
Moreover, the wave functions are related as follows
\begin{equation}
\begin{aligned}
\psi_q(t,z)=&z\frac{q(t-[z^{-1}])}{q(t)}\psi_W(t,z),\qquad
\psi_q^*(t,z)=&z^{-1}\frac{q(t+[z^{-1}])}{q(t)}\psi^*_W(t,z),\\
\psi_r(t,z)=&z^{-1}\frac{r(t-[z^{-1}])}{r(t)}\psi_W(t,z),\quad
\psi_r^*(t,z)=&z\frac{r(t+[z^{-1}])}{r(t)}\psi^*_W(t,z).
\end{aligned}
\end{equation}
\end{theorem}

\section{Squared Eigenfunction Potentials}
Let $q$ be an eigenfunction and $r$ be an adjoint eigenfunction
of the KP Lax operator $L$. For such a pair of (adjoint)
eigenfunctions Oevel \cite{Oe1}  showed that there exists a function
$\Omega(r,q)$
called the {\it squared eigenfunction potential}, which satisfies
\begin{equation}
\label{S1}
\frac{\partial \Omega(r,q)}{\partial t_k}
=Res_\partial(\partial^{-1}r(L^k)_+q\partial^{-1}).
\end{equation}
This potential was described in \cite{MS} and \cite{C} (see also
\cite{Oe1} and \cite{Oe2}). A special case of (\ref{S1}) is the case that
$k=1$,
i.e. $t_1=x$, then
\[
\frac{\partial \Omega(r,q)}{\partial x}=qr.
\]
Equation (\ref{S1}) determines $\Omega$ uniquely upto a shift by a
constant.

The expressions $\partial^{-1}r\psi$ and $\partial^{-1}\psi^*q$ are
special squared eigenfunction potentials. Notice that they are uniquely
defined.
We now want to find an explicit expression for these specific
potentials in the Segal-Wilson setting.
Since
\[
\psi_r=r^{-1}\partial^{-1}r\psi_W\quad\mbox{and}\quad
\psi_q^*=-q^{-1}\partial^{-1}q\psi_W,
\]
one has
\[
\begin{aligned}
\partial^{-1}r(t)\psi_W(t,z)&=r(t)\psi_r(t,z)\\
&=r(t)z^{-1}\frac{r(t-[z^{-1}])}{r(t)}\psi_W(t,z)\\
&=z^{-1}r(t-[z^{-1}])\psi_W(t,z)
\end{aligned}
\]
and
\[
\begin{aligned}
\partial^{-1}q(t)\psi_W^*(t,z)&=-q(t)\psi_q^*(t,z)\\
&=-q(t)z^{-1}\frac{q(t+[z^{-1}])}{q(t)}\psi_W^*(t,z)\\
&=-z^{-1}q(t+[z^{-1}])\psi_W^*(t,z).
\end{aligned}
\]
We can also express them in $s(z)$. Let
\[
q(t)=<\psi_W(t,z)|s(z)>,\quad\mbox{and}\quad
r(t)=<\psi_W^*(t,z)|\overline{zs(z)}>,
\]
then  $\partial^{-1}r\psi$ and $\partial^{-1}\psi^*q$ are equal to
\begin{equation}
\label{W0}
\begin{aligned}
S(\psi_W(t,z),\overline{zs(z)}):=&
z^{-1}<\psi_W^*(t-[z^{-1}],y)|\overline{ys(y)}>\psi_W(t,z),\\
S(\psi^*_W(t,z),s(z)):=&-z^{-1}\psi_W^*(t,z)<\psi_W(t+[z^{-1}],y)|s(y)>,
\end{aligned}
\end{equation}
respectively.

Next let $W\in Gr^{(\ell)}(H)$, then, since $\psi_W(t,z)$ and $
\psi_W^*(t,z)$ satisfy (\ref{D1}) we find the following expressions
for these squared eigenfunction potentials (see also \cite{ANP3}).
\begin{proposition}
\label{p5}
Let $|z|>|y|$, then for $W\in Gr^{(\ell)}(H)$ on has
\begin{equation}
\begin{aligned}
S(\psi_W(t,z),\overline{zs(z)})=&\frac{<(z/y)^\ell
X(t,z,y)\tau_W(t)|\overline{ys(y)}>}
{\tau_W(t)},\\
S(\psi_W^*(t,z),s(z))=&\frac{<(y/z)^\ell
X(t,y,z)\tau_W(t)|s(y)>}{\tau_W(t)},
\end{aligned}
\end{equation}
where
\begin{equation}
X(t,\lambda,\mu)=\frac{1}{\lambda-\mu}\exp(\sum t_i(\lambda^i-\mu^i))
\exp(\sum \frac{\mu^{-i}-\lambda^{-i}}{i}\frac{\partial}{\partial t_i})
\end{equation}
is the vertex operator for the KP hierarchy.
\end{proposition}
This vertex operator expresses the action of the Lie algebra
$gl_\infty$,
see \cite{KvdL} for more details.

The results of this section, will be used in the following section on
geometric B\"acklund--Darboux transformations.

\section{Geometric
B\"acklund--Darboux Transformations}
We can of course take the converse of the
procedure of section 5  and use the geometry to construct elementary
B\"acklund--Darboux transformations, viz as follows:
\begin{theorem}
\label{t2}
Let $W\in Gr(H)$ with $\tau_W$ its tau-function, $\psi_W$ its wavefunction and
$\psi_W^*$ its
adjoint wavefunction. Let $s(z)\in H$, then
$q(t)=<\psi_W(t,z)|s(z)>$ is an eigenfunction and
$r(t)=<s(z)|\overline{z\psi_W^*(t,z)}>$ is an adjoint eigenfunction
and the corresponding elementary B\"acklund--Darboux transformations
$q\partial q^{-1}$ and $r^{-1}\partial^{-1}r$ give new tau- and
wavefunctions
$\tau_{W_q}(t)=q(t)\tau_W(t)$, $\tau_{W_r}(t)=r(t)\tau_W(t)$,
$\psi_q=q\partial q^{-1}\psi_W$, $\psi_r=r^{-1}\partial^{-1}r\psi_W$
and adjoint wave functions $\psi_q^*=-q^{-1}\partial^{-1} q\psi_W^*$,
$\psi_r^*=-r\partial r^{-1}\psi_W^*$
corresponding to planes
$W_q=\{ w\in W|<w|s(z)>=0\}$, $W_r=W\oplus  \mathbb{C}s(z)$,
$W_q^\perp=W^\perp\oplus \mathbb{C}s(z)$,
$W_r^\perp=\{w\in W^\perp|<s(z)|w>=0\}$,
respectively.
\end{theorem}
We now want to generalize the situation of the previous Theorem.
Assume that is given the plane $W\in Gr^{(\ell)}(H)$ with wavefunction
$\psi_W(t,z)$ and adjoint wavefunction $\psi_W^*(t,z)$. We want to
determine the B\"acklund--Darboux transformation $BD(V,W)$ that maps
$\psi_W(t,z)$ to $\psi_V(t,z)$. We restrict ourselves to the case that
$BD(V,W)$ is a finite product of elementary B\"acklund--Darboux
transformation, i.e., to the case that $V\cap W$ has finite
codimension inside both $V$ and $W$. Of course one has that if
$\psi_W(t,z)
=P_W(\partial) e^{\sum t_iz^i}$ and $\psi_V(t,z)=P_V(\partial) e^{\sum
t_iz^i}$, that $BD(V,W)= P_V(\partial)P_W(\partial)^{-1}$. But we
assume that the wavefunction and its adjoint are only given for $W$
and not yet known for $V$.

We first start with two special cases of planes $V$ and $W$, viz.,
$V\subset W$ and $W\subset V$. First consider the former case and
assume that $V$ has codimension $n$ inside $W$. So we can find $n$
independent functions in $W\cap V^\perp$, say $w_1,w_2,\ldots,w_n$. It
is straightforward  that $BD(V,W)$ is the product of n elementary
B\"acklund--Darboux transformations of the form $q_j\partial
q_j^{-1}$.
Hence, $BD(V,W)=\sum_{k=0}^n a_k\partial^k$ with $a_k\in R$ and $a_n=1$,
i.e. is a differential operator of order $n$ with leading coefficient
equal to 1. Thus, $\psi_V=\sum_{k=0}^n
a_k\frac{\partial^k\psi_W}{\partial x^k}$ and since $<\psi_V|w_j>=0$
for all $j=1,2,\ldots, n$, we find that
\[
\sum_{k=0}^{n-1} a_k<\frac{\partial^k\psi_W(t,z)}{\partial x^k}|w_j(z)>=
-<\frac{\partial^n\psi_W(t,z)}{\partial x^n}|w_j(z)>,\quad
\mbox{for }j=1,2,\ldots,n.
\]
Using Cramer's rule one thus finds the following expressions for the
$a_k$'s:
\[
a_k=(-)^{n-k}{\cal W}(\psi_W;w_1,w_2,\ldots,w_n)^{-1}{\cal W}_k
(\psi_W;w_1,w_2,\ldots,w_n),
\]
where
\begin{equation}
{\cal W}(A;w_1,w_2,\ldots,w_n)=\det
\left (
\begin{array}{cccc}
<A|w_1>&<\frac{\partial A}{\partial x}|w_1>&\cdots
&<\frac{\partial^{n-1} A}{\partial x^{n-1}}|w_1>\\
\\
<A|w_2>&<\frac{\partial A}{\partial x}|w_2>&\cdots
&<\frac{\partial^{n-1} A}{\partial x^{n-1}}|w_2>\\
\vdots&\vdots&\ddots &\vdots\\
<A|w_n>&<\frac{\partial A}{\partial x}|w_n>&\cdots
&<\frac{\partial^{n-1}A}{\partial x^{n-1}}|w_n>
\end{array}
\right )
\end{equation}
is a Wronskian determinant and
\[
\begin{aligned}
{\cal W}_k&(A;w_1,w_2,\ldots,w_n)=\\
&\det
\left (
\begin{array}{cccccc}
<A|w_1>&\cdots&<\frac{\partial^{k-1}A}{\partial
x^{k-1}}|w_1>&
<\frac{\partial^{k+1}A}{\partial x^{k+1}}|w_1>&\cdots
&<\frac{\partial^{n}A}{\partial x^{n}}|w_1>\\
\\
<A|w_2>&\cdots&<\frac{\partial^{k-1}A}{\partial
x^{k-1}}|w_2>&
<\frac{\partial^{k+1}A}{\partial x^{k+1}}|w_2>&\cdots
&<\frac{\partial^{n}A}{\partial x^{n}}|w_2>\\
\vdots&\ &\vdots&\vdots&\ &\vdots\\
<A|w_n>&\cdots&<\frac{\partial^{k-1}A}{\partial
x^{k-1}}|w_n>&
<\frac{\partial^{k+1}A}{\partial x^{k+1}}|w_n>&\cdots
&<\frac{\partial^{n}A}{\partial x^{n}}|w_n>
\end{array}
\right ).\\
\end{aligned}
\]
This is in fact a generalization of a result of Crumm \cite{Cr}. Following
\cite{Oe1}, we give a short notation for this B\"acklund-Darboux
transformation $BD(V,W)$, viz.,
\begin{equation}
BD(V,W)={\cal W}(\psi_W;w_1,w_2,\ldots,w_n)^{-1}
{\cal W}(\psi_W;w_1,w_2,\ldots,w_n;\partial),
\end{equation}
where
\begin{equation}
\label{W1}
\begin{aligned}
{\cal W}&(A;w_1,w_2,\ldots,w_n;\partial)=\\
&\det
\left (
\begin{array}{cccc}
<A|w_1>&<\frac{\partial A}{\partial x}|w_1>&\cdots
&<\frac{\partial^{n}A}{\partial x^{n}}|w_1>\\
\\
<A|w_2>&<\frac{\partial A}{\partial x}|w_2>&\cdots
&<\frac{\partial^{n}A}{\partial x^{n}}|w_2>\\
\vdots&\vdots&\ &\vdots\\
<A|w_n>&<\frac{\partial A}{\partial x}|w_n>&\cdots
&<\frac{\partial^{n}A}{\partial x^{n}}|w_n>\\
\\
1&\partial&\cdots&\partial^n
\end{array}
\right ).
\end{aligned}
\end{equation}
The determinant (\ref{W1}) must be seen as a formal expansion with
respect to the last row vector, collecting all minors left of the
differential symbols $\partial^k$.

It is obvious that the B\"acklund-Darboux
transformation $BD(V,W)$ does not depend on the choice of basis of the
linear space $W\cap V^\perp$. It is also obvious that one can even add
elements of $W^\perp$ to the $w_k$'s.

We now want to determine the inverse of this B\"acklund-Darboux
transformation $BD(V,W)$. We immitate the proof of the Lemma in
\cite{OS}.
Let $b_k$, $1\le k\le n$ be the solution of the linear system
\[
{\cal W}(\psi_W;w_1,w_2,\ldots,w_n)b_k=\delta_{kn},
\]
i.e.,
\[
b_k=(-)^{n-k}{\cal W}(\psi_W;w_1,w_2,\ldots,w_n)^{-1}{\cal W}
(\psi_W;w_1,w_2,\ldots, w_{k-1},w_{k+1}\ldots,w_n)
\]
then
\[
BD(V,W)^{-1}=\sum_{k=1}^n<\psi_W|w_k>\partial^{-1}b_k.
\]
To show this we use the fact that for $f\in R$
\[
f\partial^{-1}=-(\partial^{-1}f)^*=\sum_{\ell=0}^\infty
\left ((-)^{\ell+1}\frac{\partial^\ell f}{\partial
x^\ell}\partial^{-\ell-1}\right )^*=
\sum_{\ell=0}^\infty \partial^{-\ell-1}\frac{\partial^\ell f}{\partial
x^\ell}
\]
and calculate
\[
\begin{aligned}
BD&(V,W)\sum_{k=1}^n<\psi_W|w_k>\partial^{-1}b_k\\&=
(BD(V,W)\sum_{k=1}^n<\psi_W|w_k>\partial^{-1}b_k)_+
+(BD(V,W)\sum_{k=1}^n<\psi_W|w_k>\partial^{-1}b_k)_-\\
&=(BD(V,W)\sum_{\ell=0}^\infty\partial^{-\ell-1}
\sum_{k=1}^n<\frac{\partial^\ell\psi_W}{\partial
x^\ell}|w_k>\partial^{-1}b_k)_+
+\sum_{k=1}^nBD(V,W)(<\psi_W|w_k>)\partial^{-1}b_k\\
&=(BD(V,W)\sum_{\ell=0}^{n-1}\partial^{-\ell-1}{\cal
W}(\psi_W;w_1,w_2,\ldots,w_n))_+ +0\\
&=(BD(V,W)\partial^{-n})_+\\
&=1.
\end{aligned}
\]

Again one has a short notation for
\begin{equation}
BD(V,W)^{-1}={\cal W}(\partial;\psi_W;w_1,w_2,\ldots,w_n)
{\cal W}(\psi_W;w_1,w_2,\ldots,w_n)^{-1},
\end{equation}
where
\begin{equation}
\label{W3}
\begin{aligned}
{\cal W}(\partial;A;&w_1,w_2,\ldots,w_n)=\\
&\det
\left (
\begin{array}{ccccc}
<A|w_1>&<\frac{\partial A}{\partial x}|w_1>&\cdots
&<\frac{\partial^{n-1} A}{\partial x^{n-2}}|w_1>
&<A|w_1>\partial^{-1}\\
\\
<A|w_2>&<\frac{\partial A}{\partial x}|w_2>&\cdots
&<\frac{\partial^{n-1} A}{\partial x^{n-2}}|w_2>
&<A|w_2>\partial^{-1}\\
\vdots&\vdots&\ &\vdots\\
<A|w_n>&<\frac{\partial A}{\partial x}|w_n>&\cdots
&<\frac{\partial^{n-1}A}{\partial x^{n-2}}|w_n>
&<A|w_n>\partial^{-1}.
\end{array}
\right )\\
\end{aligned}
\end{equation}
In (\ref{W3}) one has to use the expansion of the determinant with
respect to the last column collecting all minors on the right of the
$<A|w_k>\partial^{-1}$ symbols.

Next consider the other special case, viz., that $W\subset V$ of
codimension $m$. Then clearly $BD(V,W)$ is the product of $m$
elementary adjoint B\"acklund-Darboux transformations
$r_i^{-1}\partial^{-1}r_i$. We first do not consider $BD(V,W)$ but
rather $BD(V,W)^{*-1}$, which is a differential operator of order $m$,
say
$\sum_{k=0}^m a_k\partial^k$. Then notice that $a_m=(-1)^m$ and
$\psi_V^*=\sum_{k=0}^m a_k\frac{\partial^k\psi_W^*}{\partial x^k}$.
Now let $v_1,v_2,\ldots,v_m$ be $m$ linearly independent elements in
$W^\perp\cap V$. For notational convenience we introduce $v^\dagger$
by
\begin{equation}
\label{W4}
v^\dagger(z):=\overline {zv(z)}.
\end{equation}
Then  $<\psi_V^*|v_j^\dagger>=0$
and hence

\[
\sum_{k=0}^{m-1}a_k<\frac{\partial^k\psi_W^*}{\partial
x^k}|v_j^\dagger>=
(-)^{m+1}<\frac{\partial^m\psi_W^*}{\partial
x^m}|v_j^\dagger>, \quad\mbox{for }j=1,2,\dots,m.
\]
Using Cramer's rule again one obtains the following expression for the
$a_k$'s:
\[
a_k=(-)^k{\cal
W}(\psi_W^*;v^\dagger_1,v^\dagger_2,\ldots,v^\dagger_m)^{-1}
{\cal
W}_k(\psi_W^*;v^\dagger_1,v^\dagger_2,\ldots,v^\dagger_m).
\]
In a similar way as before one finds
\begin{equation}
BD(V,W)^{*-1}=(-)^m{\cal
W}(\psi_W^*;v^\dagger_1,v^\dagger_2,\ldots,v^\dagger_m)^{-1}
{\cal
W}(\psi_W^*;v^\dagger_1,v^\dagger_2,\ldots,v^\dagger_m;\partial),
\end{equation}
where one has to use the expansion of ${\cal
W}(\psi_W^*;v^\dagger_1,v^\dagger_2,\ldots,v^\dagger_m;\partial)$
with respect to the last row collecting all minors on the left of the
$\partial$ symbols.
So
\begin{equation}
\label{W5}
BD(V,W)^{-1}=(-)^m{\cal
W}(\psi_W^*;v^\dagger_1,v^\dagger_2,\ldots,v^\dagger_m;-\partial){\cal
W}(\psi_W^*;v^\dagger_1,v^\dagger_2,\ldots,v^\dagger_m)^{-1}
\end{equation}
and here one has to use the expansion of ${\cal
W}(\psi_W^*;v^\dagger_1,v^\dagger_2,\ldots,v^\dagger_m;-\partial)$
with respect to the last row collecting all minors on the right of the
$\partial$ symbols.

In a similar way as before we now determine
$BD(V,W)^{*}$. Let $b_k$, $1\le k\le m$ be a solution of the linear
system
\[
{\cal
W}(\psi_W^*;v^\dagger_1,v^\dagger_2,\ldots,v^\dagger_m)b_k=(-1)^m\delta_{mk},
\]
then
\[
BD(V,W)^{*}=\sum_{k=1}^m <\psi_W^*|v_k^\dagger >\partial^{-1}b_k,
\]
with
\[
b_k=(-)^k
{\cal
W}(\psi_W^*;v^\dagger_1,v^\dagger_2,\ldots,v^\dagger_m)^{-1}
{\cal
W}(\psi_W^*;v^\dagger_1,\ldots,v^\dagger_{k-1},
v^\dagger_{k+1},\ldots,v^\dagger_m).
\]
So
\[
BD(V,W)=
-\sum_{k=1}^mb_k\partial^{-1} <\psi_W^*|v_k^\dagger >
\]
and hence
\begin{equation}
\label{W6}
BD(V,W)=(-)^m{\cal
W}(\psi_W^*;v^\dagger_1,v^\dagger_2,\ldots,v^\dagger_m)^{-1}
{\cal
W}(\partial;\psi_W^*;v^\dagger_1,v^\dagger_2,\ldots,v^\dagger_m)^*,
\end{equation}
where one has to use the expansion of the determinant of
${\cal
W}(\partial;\psi_W^*;v^\dagger_1,v^\dagger_2,\ldots,v^\dagger_m;)$ with
respect to the last column collecting all minors on the right of the
$<\psi_W^*|v_k^\dagger>\partial^{-1}$ symbols.
Again also in this case it is clear that everything is independent
from the choice of basis of $W^\perp\cap V$.

We now use the results of the previous two cases to determine the
B\"acklund-Darboux transformation $BD(V,W)$ in the more general case
that $V\cap W$ has codimension $m$ in $V$ and codimension $n$ in $W$.
Let $w_1,w_2,\ldots,w_n$ be a basis of $V^\perp\cap W$
 and let $v_1,v_2,\ldots,v_m$ be a basis of $W^\perp\cap V$.
Starting with $W$ we first apply $m$ elementary adjoint
B\"acklund-Darboux transformations to obtain $W+V$ and then apply $n$
elementary
B\"acklund-Darboux transformations to obtain $V$. In other words
\[
BD(V,W)=BD(V,V+W)BD(V+W,W).
\]
Now $BD(V,V+W)$ is a differential operator of order n for which the
leading coefficient is equal to $1$ and
$BD(V+W,W)=\sum_{k=1}^mb_k\partial^{-1}<\psi^*_W|v_k^\dagger>$and is in
fact a pseudodifferential operator of order $-m$, again with leading
coefficient equal to $1$. Hence
\[
\begin{aligned}
BD(V,W)&=\left (\sum_{\ell=0}^nc_\ell \partial^\ell\right )
\left (\sum_{k=1}^mb_k\partial^{-1}<\psi^*_W|v_k^\dagger>\right )\\
&=\sum_{k=1}^ma_{-k}\partial^{-1}<\psi^*_W|v_k^\dagger>+\sum_{\ell=0}^{n-m}
a_\ell\partial^\ell
\end{aligned}
\]
is a pseudodifferential operator of order $n-m$ with leading
coefficient equal to $1$ (notice that this is $a_{n-m}$ if $ n\ge m$).
Since
$BD(V,W)\psi_W$ belongs to $V$ it must be perpendicular to
$w_1,w_2,\ldots,w_n$, which gives the restriction
\begin{equation}
\label{W7}
\sum_{k=1}^ma_{-k}S(\psi_W,v_k^\dagger,w_j)+\sum_{\ell=0}^{n-m}
a_\ell<\frac{\partial^\ell\psi_W}{\partial x^\ell}|w_j>=0,
\end{equation}
Where (cf. (\ref{W0}))
\begin{equation}
\label{W8}
S(A,s,t):=<S(A,s)|t>.
\end{equation}
Now if $n\ge m$, the equations (\ref{W7}) determine $BD(V,W)$, since
$a_{n-m}=1$.
However, if $n<m$ then (\ref{W7}) is equal to
\begin{equation}
\label{W9}
\sum_{k=1}^ma_{-k}S(\psi_W,v_k^\dagger,w_j)=0,
\end{equation}
and we need some more information. We also know that $BD(V,W)$ is a
pseudodifferential operator of order $n-m$ with leading coefficient
equal to 1. This leads to the extra restrictions.
\begin{equation}
\label{W10}
\sum_{k=1}^ma_{-k}<\frac{\partial^{j}\psi_W^*}{\partial
x^j}|v_k^\dagger>
=(-)^{m-n-1}\delta_{j,m-n-1} \quad\mbox{for }j=0,1,\ldots m-n-1.
\end{equation}
Then (\ref{W9}) and (\ref{W10}) determine $BD(V,W)$ uniquely.
Using Cramer's rule we can determine in the case that $n\ge m$ from
(\ref{W7}) and in the case that $m>n$ from (\ref{W9}) and (\ref{W10})
the coefficients $a_i$ of $BD(V,W)$.
We thus obtain
\begin{equation}
\label{W11}
BD(V,W)={\cal W}(\psi_W;w_1,\dots ,w_n;\psi^*_W;v_1^\dagger
,\ldots,v_m^\dagger)^{-1}
{\cal W}(\psi_W;w_1,\dots
,w_n;\psi^*_W;v_1^\dagger,\ldots,v_m^\dagger;\partial),
\end{equation}
where
\begin{equation}
\label{W12}
\begin{aligned}
{\cal W}&(A;w_1,\dots ,w_n;B;v_1,\ldots,v_m)=\\
&\det
\left (
\begin{array}{cccccc}
S(A,v_1,w_1)&\cdots&S(A,v_m,w_1)&
<A|w_1>
&\cdots
&<\frac{\partial^{n-m-1}A}{\partial x^{n-m-1}}|w_1>\\
\vdots&\ &\vdots&\vdots&\ &\vdots\\
S(A,v_1,w_n)&\cdots&S(A,v_m,w_n)&
<A|w_n>
&\cdots
&<\frac{\partial^{n-m-1}A}{\partial x^{n-m-1}}|w_n>\\
\\
<B|v_1>&\cdots&<B|v_m>&&&\\
\\
<\frac{\partial B}{\partial x}|v_1>&\cdots&<\frac{\partial B}{\partial
x}|v_m>&&&\\
\vdots&\ &\vdots&&&\\
<\frac{\partial^{m-n-1} B}{\partial x^{m-n-1}}|v_1>
&\cdots&<\frac{\partial^{m-n-1} B}{\partial x^{m-n-1}}|v_m>&&&\\
\end{array}
\right )
\end{aligned}
\end{equation}
and
\begin{equation}
\label{W13}
\begin{aligned}
{\cal W}&(A;w_1,\dots ,w_n;B;v_1,\ldots,v_m;\partial)=\\
&\det
\left (
\begin{array}{cccccc}
S(A,v_1,w_1)&\cdots&S(A,v_m,w_1)&
<A|w_1>
&\cdots
&<\frac{\partial^{n-m}A}{\partial x^{n-m}}|w_1>\\
\vdots&\ &\vdots&\vdots&\ &\vdots\\
S(A,v_1,w_n)&\cdots&S(A,v_m,w_n)&
<A|w_n>
&\cdots
&<\frac{\partial^{n-m}A}{\partial x^{n-m}}|w_n>\\
\\
\partial^{-1}<B|v_1>&\cdots&\partial^{-1}<B|v_m>&1&\ldots&\partial^{n-m}\\
\\
<B|v_1>&\cdots&<B|v_m>&&&\\
\\
<\frac{\partial B}{\partial x}|v_1>&\cdots&<\frac{\partial B}{\partial
x}|v_m>&&&\\
\vdots&\ &\vdots&&&\\
<\frac{\partial^{m-n-2} B}{\partial x^{m-n-2}}|v_1>
&\cdots&<\frac{\partial^{m-n-2} B}{\partial x^{m-n-2}}|v_m>&&&\\
\end{array}
\right ).
\end{aligned}
\end{equation}
The matrix appearing in (\ref{W12}) is an $n\times n$-matrix if $n\ge
m$ and an $m\times m$-matrix if $n<m$.
The matrix appearing in (\ref{W13}) is an $n+1\times n+1$-matrix if $n\ge
m$ and an $m\times m$-matrix if $n<m$. The determinant appearing in
(\ref{W13}) is the formal expansion with respect to the $(n+1)$-th
row putting all the minors to the left of the $\partial$ symbols.

Finally we want to calculate the inverse of $BD(V,W)$ or rather the
adjoint of the inverse. First notice that
\begin{equation}
\begin{aligned}
\label{W14}
BD(V,W)^{*-1}&=BD(V^\perp, W^\perp)\quad\mbox{and}\\
BD(V,W)^{*-1}&=BD(V,V\cap W^{*-1}BD(V\cap W,W)^{*-1}.
\end{aligned}
\end{equation}
{}From the first line of (\ref{W14}) we deduce that
\[
<BD(V,W)^{*-1}\psi_W^*|v_j^\dagger>=0\quad\mbox{for }j=1,2,\ldots,m
\]
and from the second line that
\[
BD(V,W)^{*-1}=\sum_{\ell=1}^na_{-\ell}\partial^{-1}<\psi_W|w_\ell>+
\sum_{k=0}^{m-n}a_k\partial^k.
\]
Using the fact that
$BD(V,W)^{*-1}$ is an $(m-n)$-th order pseudodifferential operator
with leading coefficient $(-1)^{m-n}$ we obtain that for $1\le j\le m$
\begin{equation}
\label{W15}
\sum_{\ell=1}^na_{-\ell}S(\psi_W^*,w_\ell,v^\dagger_j)+
\sum_{k=0}^{m-n-1}a_k<\frac{\partial^k \psi_W^*}{\partial
x^k}|v^\dagger_j>=(-1)^{n-m-1},
\end{equation}
if $m\ge n$ and that if $m<n$
\begin{equation}
\label{W16}
\begin{aligned}
\sum_{\ell=1}^na_{-\ell}S(\psi_W^*,w_\ell,v^\dagger_j)=&0,\\
\sum_{\ell=1}^na_{-\ell}<\frac{\partial^k\psi_W}{\partial x^k}|w_\ell>
=&-\delta_{k,n-m-1},
\end{aligned}
\end{equation}
for $1\le j\le m$ and $0\le k\le n-m-1$.
This determines
$BD(V,W)^{*-1}$ uniquely.
One has
\begin{equation}
\begin{aligned}
\ &BD(V,W)^{*-1}\\
 &=(-)^{n+m}{\cal W}
(\psi_W^*;v_1^\dagger,\ldots,v_m^\dagger;\psi_W;w_1,\ldots,w_n)^{-1}
{\cal W}
(\psi_W^*;v_1^\dagger,\ldots,v_m^\dagger;\psi_W;w_1,\ldots,w_n;\partial),
\end{aligned}
\end{equation}
hence the adjoint of this operator is
\begin{equation}
\begin{aligned}
\ &BD(V,W)^{-1}=\\
 &(-)^{n+m}{\cal
W}(\psi_W^*;v_1^\dagger,\ldots,v_m^\dagger;\psi_W;w_1,\ldots,w_n;\partial)^*
{\cal
W}(\psi_W^*;v_1^\dagger,\ldots,v_m^\dagger;\psi_W;w_1,\ldots,w_n)^{-1}
,
\end{aligned}
\end{equation}

We now want to calculate $\tau_V$. In order to do that we write
$BD(V,W)$ as product of $n+m$ elementary B\"acklund-Darboux
transformations
\[
BD(V,W)=BD(V,U_{n+m-1})BD(U_{n+m-1},U_{n+m-2})\cdots BD(U_1,W),
\]
where
\[
U_k=
\begin{cases}
W+\sum_{j=1}^k \mathbb{C}v_j&\mbox{for }1\le k\le m,\\
\{u\in W+V|<u|w_j>=0\ \mbox{for }1\le j\le k-n\}&\mbox{for }m\le k\le m+n.
\end{cases}
\]
Then from Theorem \ref{t2} we deduce that for $k\le m$
\[
\begin{aligned}
\tau_{U_k}&=<\psi_{U_{k-1}}^*|v_k^\dagger>\tau_{U_{k-1}}\\
&=<BD(U_{k-1},W)^{*-1}\psi_W^*|v_k^\dagger>\tau_{U_{k-1}}\\
&=(-)^{k-1}\frac{{\cal W}(\psi^*_W;v_1^\dagger,\ldots,v_{k}^\dagger)}
{{\cal W}(\psi^*_W;v_1^\dagger,\ldots,v_{k-1}^\dagger)}\tau_{U_{k-1}}
\end{aligned}
\]
and that for $k>m$
\[
\begin{aligned}
\tau_{U_k}&=<\psi_{U_{k-1}}|w_{k-m}>\tau_{U_{k-1}}\\
&=<BD(U_{k-1},W)\psi_W |w_{k-m}>\tau_{U_{k-1}}\\
&=\frac{{\cal W}(\psi_W;w_1,\ldots ,w_{k-m};\psi^*_W;
v_1^\dagger,\ldots,v_{m}^\dagger)}
{{\cal W}(\psi_W;w_1,\ldots ,w_{k-m-1};
\psi^*_W;v_1^\dagger,\ldots,v_{m}^\dagger)}\tau_{U_{k-1}}.
\end{aligned}
\]
{}From which we deduce that (upto a scalar multiple)
\begin{equation}
\tau_V={\cal W}(\psi_W;w_1,\ldots ,w_n;
\psi^*_W;v_1^\dagger,\ldots,v_{m}^\dagger)\tau_W.
\end{equation}
Notice that all the above considerations are independent of the choice
of bases of $V\cap W^\perp$ and $V^\perp\cap W$. Thus we have proven
the main theorem of this paper
\begin{theorem}
\label{t3}
Let $V\in Gr^{(k)}$ and $W\in Gr^{(\ell)}$ be such that $V\cap W$ has
codimension $m$ in $V$ and codimension $n$ in $W$ (hence $k-m=\ell-n$)
and let $\psi_W$, $\psi_W^*$ and $\tau_W$ be the to $W$ corresponding
wave-, adjoint wave- and tau-function. Then the corresponding
B\"acklund-Darboux transformation is equal to (cf. (\ref{W12}), (\ref{W13}))
\[
BD(V,W)={\cal W}(\psi_W;w_1,\dots ,w_n;\psi^*_W;v_1^\dagger
,\ldots,v_m^\dagger)^{-1}
{\cal W}(\psi_W;w_1,\dots
,w_n;\psi^*_W;v_1^\dagger,\ldots,v_m^\dagger;\partial),
\]
and its inverse is equal to
\[
\begin{aligned}
\  &BD(V,W)^{-1}=\\
 &(-)^{n+m}{\cal
W}(\psi_W^*;v_1^\dagger,\ldots,v_m^\dagger;\psi_W;w_1,\ldots,w_n;\partial)^*
{\cal
W}(\psi_W^*;v_1^\dagger,\ldots,v_m^\dagger;\psi_W;w_1,\ldots,w_n)^{-1}
\end{aligned}
\]
for any bases $w_1,w_2,\dots,w_n$ of $W\cap V^\perp$ and
$v_1,v_2,\ldots,v_m$
of $V\cap W^\perp$.
Moreover,
\[
\begin{aligned}
\psi_V&=
{\cal W}(\psi_W;w_1,\dots ,w_n;\psi^*_W;v_1^\dagger ,\ldots,v_m^\dagger)^{-1}
{\cal W}(\psi_W;w_1,\dots
,w_n;\psi^*_W;v_1^\dagger,\ldots,v_m^\dagger;\psi_W),\\
\psi_V^*&=\\
&(-)^{n+m}{\cal
W}(\psi_W^*;v_1^\dagger,\ldots,v_m^\dagger;\psi_W;w_1,\ldots,w_n)^{-1}{\cal
W}(\psi_W^*;v_1^\dagger,\ldots,v_m^\dagger;\psi_W;w_1,\ldots,w_n;\psi_W^*),
\end{aligned}
\]
and
\[
\tau_V={\cal W}(\psi_W;w_1,\ldots ,w_n;
\psi^*_W;v_1^\dagger,\ldots,v_{m}^\dagger)\tau_W,
\]
where
\begin{equation}
\label{5.1}
\begin{aligned}
{\cal W}&(A;w_1,\dots ,w_n;B;v_1,\ldots,v_m;C)=\\
&\det
\left (
\begin{array}{cccccc}
S(A,v_1,w_1)&\cdots&S(A,v_m,w_1)&
<A|w_1>
&\cdots
&<\frac{\partial^{n-m}A}{\partial x^{n-m}}|w_1>\\
\vdots&\ &\vdots&\vdots&\ &\vdots\\
S(A,v_1,w_n)&\cdots&S(A,v_m,w_n)&
<A|w_n>
&\cdots
&<\frac{\partial^{n-m}A}{\partial x^{n-m}}|w_n>\\
\\
S(C,v_1)&\cdots&S(C,v_m)&C&\ldots&\frac{\partial^{n-m} C}{\partial x^{n-m}}\\
\\
<B|v_1>&\cdots&<B|v_m>&&&\\
\\
&\cdots&<\frac{\partial B}{\partial x}|v_m>&&&\\
\vdots&\ &\vdots&&&\\
<\frac{\partial^{m-n-2} B}{\partial x^{m-n-2}}|v_1>
&\cdots&<\frac{\partial^{m-n-2} B}{\partial x^{m-n-2}}|v_m>&&&\\
\end{array}
\right ),
\end{aligned}
\end{equation}
where
$S(C,v)=\partial^{-1}(<B|v>C)$ is defined in (\ref{W0})and $S(A,v,w)$ in
(\ref{W8}).
\end{theorem}

\section{B\"acklund--Darboux Transformations for the Gelfand--Dickey Hierarchy}

In this section we want to consider the elementary
B\"acklund--Darboux transformations
for the Gelfand--Dickey hierarchies. It is well-known that these correspond
to the $n$-th reduced KP hierarchy, i.e., $L^n=(L^n)_+$. Since
$L^n\psi_W(t,z)=z^n\psi_W(t,z)$ we have a trivial B\"acklund--Darboux
transformation
that maps $W$ into $z^nW$. Since $L^n=(L^n)_+$,
$z^n\psi_W(t,z)=\partial_n( \psi_W(t,z))$ and hence $z^nW\subset W$.
We now want to determine which vectors $s\in W$ we can choose
such that
$W_s=\{ w\in W\, <w|s>=0\}$
also satisfies $z^nW_s\subset W_s$.

We will first prove the following Lemma.
\begin{lemma}
\label{GDl1}
Let $W'\subset W$ such that $z^nW'\subset W$ and
\[
W_s=\{ w\in W|<w|s>=0\}\quad\hbox{and }W'_s=\{ w\in W'| <w|s>=0\}.
\]
Then $z^nW'_s\subset W_s$ if and only if there exists a $\lambda\in\mathbb{C}$
such that
\begin{equation}
\label{GD1}
(z^{-n}-\lambda)s\in{W'}^\perp.
\end{equation}
\end{lemma}
\noindent
{\bf Proof.}
Let $v\in W'_s$ and $z^nv\in W_s$, then
\[
<z^nv|s(z)>=<v|z^{-n}s(z)>=0,
\]
in other words $z^{-n}s(z)\in {W'_s}^\perp$. Hence, since
${W'_s}^\perp= {W'}^\perp\oplus\mathbb{C}s$, there must exist a
$\lambda\in\mathbb{C}$ such that $(z^{-n}-\lambda)s\in{W'}^\perp$.

Now assume that (\ref{GD1}) holds, then for every $v\in W'_s$ one has
\[
<z^nv|s(z)>=<v|z^{-n}s(z)>=\overline{\lambda}<v|s(z)>=0,
\]
hence $z^nv\in W_s$.\hfill$\square$

\noindent
So it suffices to assume that $W=W'$ and that $s(z)\in W$ satisfies
(\ref{GD1}). Let
\begin{equation}
\label{GD2}
U=W\cap (z^nW)^\perp ,
\end{equation}
then (see e.g. \cite{PS})
$W=\bigoplus_{k=0}^\infty z^{kn}U$,
so we write
\[
s(z)=\sum_{k=0}^\infty u_kz^{kn}\quad\hbox{with }u_k\in U.
\]
Since $(z^{-n}-\lambda)s(z)\in W^\perp$, one obtains that
$u_{k+1}=\lambda u_k$ for $k=0,1,\ldots$ and hence,
\[
s(z)=\sum_{k=0}^\infty(\lambda z)^{kn}u_0\quad\hbox{with }u_0\in U\hbox{ and
}\lambda\in\mathbb{C}.
\]
Since
\[
\begin{aligned}
<s(z)|s(z)>&=<\sum_{k=0}^\infty(\lambda z)^{kn}u_0|
\sum_{k=0}^\infty(\lambda z)^{kn}u_0>\\
\ &=\sum_{k=0}^\infty
|\lambda|^{2k}|u_0|^2,\\
\end{aligned}
\]
must be finite, we must choose $|\lambda|<1$.
Notice that if $W$ satisfies $z^nW\subset W$, then $W^\perp$ satisfies
$z^{-n}W^\perp\subset W^\perp$.
Now let $V=W^\perp\cap(z^{-n}W^\perp)^\perp$, then $V=z^{-n}U$ and hence
one obtains that $t(z)\in W^\perp$ which gives the correct
B\"acklund--Darboux transformation is
\[
t(z)=\sum_{k=0}^\infty (\frac{\lambda}{z})^{nk}z^{-n}u_0,
\quad \hbox{with }u_0\in U\hbox{ and }\lambda\in\mathbb{C},|\lambda|<1.
\]
We thus have proven the following theorem
\begin{theorem}
\label{GDt1}
Let $W\in Gr(H)$ satisfy $z^nW\subset W$
and hence $z^{-n}W^\perp\subset W^\perp$,
then $s(z)\in W$ (resp. $t(z)\in W^\perp$)
defines an eigenfunction $q(t)=<\psi_w(t,z)|s(z)>$
(resp. adjoined eigenfunction $r(t)=<s(z)|\overline{z\psi_W^*(t,z)}>$)
and $\psi_q(t,z)=q\partial q^{-1}\psi_W(t,z)$ (resp.
$\psi_r(t,z)=r^{-1}\partial^{-1}r\psi_W(t,z)$) be a new wavefunction
defined by the elementary B\"acklund-Darboux
transformation that maps $W$
into $W_q=\{w\in W|<w|s(z)>=0\}$ (resp. $W_r=W\oplus\mathbb{C}t(z)$).
Let $U=W\cap (z^nW)^\perp$. Then $W_q$ satisfies $z^nW_q\subset W_q$
(resp.$ W_r$ satisfies $z^nW_r\subset W_r$) if and only if
\[
s(z)=\sum_{k=0}^\infty(\lambda z)^{kn}u_0\quad(\hbox{resp. }
t(z)=\sum_{k=0}^\infty (\frac{\lambda}{z})^{nk}z^{-n}u_0),
\]
with $u_0\in U$ and $\lambda\in\mathbb{C}$, $|\lambda|<1$.
\end{theorem}

Next we want to give the geometrical interpretation (in terms of the
Segal--Wilson Grassmannian) of
the classical Darboux transformation of an $n$-th Gelfand--Dickey
Lax operator $L^n$.

The classical Darboux transformation consists of
factorizing an $n$-th order differential operator
$Q=RS$, with $R$ and $S$ differential operators of order $r$, $s$,
respectively, with $r+s=n$, and then exchanging
the place of the factors, i.e.,  defining a new differential operator
$Q'=SR$. We assume that $Q=L^n$ is a Gelfand--Dickey operator
and we want that $Q'$ is again a Gelfand--Dickey operator. Let
$L=P_W\partial P_W^{-1}$, with $W\in Gr(H)$, since
$Q'=SQS^{-1}$, we find that  $Q'={L'}^n$, with $L'=P_{W'}\partial
P_{W'}^{-1}$ and $P_{W'}=SP_{W}$. Because $Q=RS$ we find that
\[
z^nW\subset W'\subset W,
\]
hence the vectors $w_1(z), w_2(z),\ldots, w_s(z)\in W$, that define the
B\"acklund--Darboux transformation $S$, must be perpendicular to $z^nW$
and thus must belong to $U$, defined by (\ref{GD2}).
We thus have proven
\begin{theorem}
\label{GDt2}
Let $W\in Gr(H)$ satisfy $z^nW\subset W$ and let $L=P_w\partial
p_W^{-1}$.
The classical Darboux
transformation of the Gelfand--Dickey differential operator $Q=L^n$,
consists of choosing $s$  ($0<s<n$) linearly independent vectors
$v_j\in U$ ($U$ defined by (\ref{GD2})), such that
\[
S={\cal W}(\psi_W;w_1,w_2,\ldots,w_s)^{-1}
{\cal W}(\psi_W;w_1,w_2,\ldots,w_s;\partial).
\]
Then $Q'=SQS^{-1}$ is again a Gelfand--Dickey differential operator.
\end{theorem}

\section{From KP to the 1-Toda Lattice Hierarchy}

The 1-Toda lattice hierarchy, as described in \cite{AvM2} (see also
\cite{AvM3} and \cite {HH})
is the set of deformation equations (Lax
equations)
\begin{equation}
\label{6.1}
\frac{\partial L}{\partial_n}=[(L^n)_+,L],
\end{equation}
for infinite matrices
\begin{equation}
\label{6.2}
L=L(t,\Lambda)=\Lambda+\sum_{i=0}^\infty
a_i\Lambda^{-i}\quad\mbox{with }
\Lambda=\sum_{k\in\mathbb{Z}}\frac{\epsilon_k}{\epsilon_{k+1}}E_{k,k+1},
\end{equation}
depending on $t=(t_1,t_2,\ldots)$ where $\partial_n=\partial/\partial
t_n$ and where
$\epsilon_k\in\mathbb{C}^\times$,
$a_i=\sum_{k\in\mathbb{Z}} a_i(k,t)E_{kk}$ is an infinite diagonal matrix.
Here $A_+$ refers to the upper-diagonal part of the matrix $A$,
inclusing the diagonal. We denote by $A_-=A-A_+$. For notational
covenience we write
$\Lambda=\epsilon\sum_{k\in\mathbb{Z}}E_{k,k+1}\epsilon^{-1}$ with
$\epsilon=\sum_{k\in\mathbb{Z}}\epsilon_kE_{kk}$.

To the problem (\ref{6.1}) one associates wavevectors $\Psi(t,z)$ and
adjoint wavevectors $\Psi^*(t,z)$,
\begin{equation}
\label{6.3}
\begin{aligned}
\Psi(t,z)&=(\epsilon_n\psi_n(t,z))_{n\in\mathbb{Z}}
=P(t,\Lambda)e^{\sum_{k=0}^\infty
t_k\Lambda^k}\epsilon\chi(z),\\
\Psi^*(t,z)&=(\epsilon_n^{-1}z\psi_{n+1}^*(t,z))_{n\in\mathbb{Z}}
=(P(t,\Lambda)^T)^{-1}e^{-\sum_{k=0}^\infty
t_k\Lambda^{-k}}\epsilon^{-1}\chi(z^{-1}),
\end{aligned}
\end{equation}
which satisfies
\begin{equation}
\label{6.4}
\begin{aligned}
L\Psi(t,z)&=&z\Psi(t,z),\quad L^T\Psi^*(t,z)&=&z\Psi^*(t,z),\\
\partial_n(\Psi(t,z))&=&(L^n)_+\Psi(t,z),\quad
\partial_n(\Psi^*(t,z))&=&((L^n)_+)^T\Psi^*(t,z).
\end{aligned}
\end{equation}
Here $A^T$ stands for the transposed of the matrix $A$,
$\chi(z)$ is the infinite column-vector $\chi(z)=(z^n)_{n\in\mathbb{Z}}$
and
\begin{equation}
\label{6.5}
P(t,\Lambda)=I+\sum_{k=1}^\infty p_k \Lambda^{-k}.
\end{equation}
Notice that in this situation
$L=P(t,\Lambda)\Lambda P(t,\Lambda)^{-1}$.
We can formulate the 1-Toda lattice hierarchy in an other way, viz. as
follows (see also \cite{AvM2}):
\begin{proposition}
\label{p6.1}
The following two statements are equivalent:

\noindent
(a) The Lax operator $L$ of the form (\ref{6.2}) satisfies the
1-Toda lattice hierarchy (\ref{6.1}).

\noindent
(b)
$\psi_n(t,z)$ is a wavefunction of the KP hierarchy for all
$n\in\mathbb{Z}$, with $\psi_n^*(t,z)$ its adjoint wavefunction. These
(adjoint) wave functions satisfy the $n-m$-th modified KP hierarchy:
\begin{equation}
\label{6.6}
<\psi_n(t,z)|\overline{z\psi_m^*(s,z)}>=0\quad \mbox{for all }n\ge m.
\end{equation}
\end{proposition}

\noindent
To prove this proposition we use the following lemma:
\begin{lemma}
\label{l6.1}
(a) Let $U$ and $V$ be two infinite matrices, then
\[
Res_{z=0}\frac{1}{z}U\chi(z)(V\chi(z^{-1}))^T=UV^T.
\]

\noindent
(b) Let $P(t,\partial)$ and $Q(t,\partial)$ be two pseudodifferential
operators such that
\[
Res_{z=0}\, P(t,z)Q(t,-z)e^{\sum (t_i-s_i)z^i}=0,
\]
then $(P(t,\partial)Q(t,\partial)^*)_-=0$.
\end{lemma}
{\bf Proof.} For a proof of (b) we refer to \cite{KvdL}. To prove (a), let
$U=\sum_{i,j}u_{ij}E_{ij}$ and $V=\sum_{i,j}v_{ij}E_{ij}$,
then
\[
\begin{aligned}
Res_{z=0}\frac{1}{z}U\chi(z)(V\chi(z^{-1}))^T=&Res_{z=0}\frac{1}{z}
\sum_{i,j,k,\ell}u_{ij}z^jv_{k\ell}z^{-\ell}E_{ik}\\
=&\sum_{i,j,k}u_{ij}v_{kj}E_{ik}\\
=&UV^T,
\end{aligned}
\]
which proves (a).\hfill$\square$
\
\\

\noindent
{\bf Proof of Proposition \ref{p6.1}.}
For all $k\ge 0$ one has
\[
\begin{aligned}
0=(\Lambda^k)_-=&(\Lambda^kP(t,\lambda)P(t,\lambda)^{-1})_-\\
=&(\Lambda^kP(t,\lambda)e^{\sum t^i\Lambda^i}
e^{-\sum t^i\Lambda^i}P(t,\lambda)^{-1})_-,
\end{aligned}
\]
hence
\[
Res_{z=0}\, (\frac{1}{z}\Lambda^k(\Psi(t,z))\Psi^*(t,z)^T)_-=0
\]
and thus
\[
\begin{aligned}
0=&Res_{z=0}\,
(\frac{1}{z}((L^{n_1})_+)^{k_1}((L^{n_2})_+)^{k_2}\ldots
((L^{n_m})_+)^{k_m}
(\Psi(t,z))\Psi^*(t,z)^T)_-\\
=&Res_{z=0}\,
(\frac{1}{z}
\frac{\partial^{k_1+\ldots+k_m} (\Psi(t,z))}
{\partial t_{n_1}^{k_1}\cdots\partial t_{n_m}^{k_m}}
\Psi^*(t,z)^T)_-.
\end{aligned}
\]
Now using Taylor's formula one obtains
\begin{equation}
\label{6.7}
Res_{z=0}\,
(\frac{1}{z}\Psi(t,z)\Psi^*(s,z)^T)_-=0,
\end{equation}
which is equivalent to (\ref{6.6}).
Next we write the oscillating functions $\psi_n(t,z)$ and
oscillating function
$\psi^*(t,z)$ in terms of pseudodifferential
operators.
\[
\psi_n(t,z)=P(\partial)\partial^ne^{\sum t_iz^i}\quad \mbox{and}\quad
\psi_n^*(t,z)=Q(\partial)(-\partial)^{-n}e^{-\sum t_iz^i}.
\]
Using Lemma \ref{l6.1}(b), one deduces from (\ref{6.6}) for $n=m$ that
$Q^*=P^{-1}$ and hence $\psi_n^*(t,z)$ is the adjoint oscillating
function of $\psi_n(t,z)$. Since they satisfy (\ref{6.6}) for $n=m$,
$\psi_n(t,z)$ (resp. $\psi_n^*(t,z)$) must be a KP wavefunction (resp.
adjoint wavefunction).

To prove the converse we use the bilinear identity (\ref{6.7}) which
is equivalent to (\ref{6.6}). Denote
\[
\Psi(t,z)=P(t,\Lambda)e^{\sum t_i\Lambda^i}\epsilon\chi(z)\quad \mbox{and}\quad
\Psi^*(t,z)=Q(t,\Lambda)e^{-\sum
t_i\Lambda^{-i}}\epsilon^{-1}\chi(z^{-1}),
\]
then using part (a) of Lemma \ref{l6.1}, we deduce from
(\ref{6.7}) with $s=t$ that
$(P(t,\Lambda)Q(t,\Lambda)^T)_-=0$
and hence
\begin{equation}
\label{6.8}
Q(t,\Lambda)^T=P(t,\Lambda)^{-1}.
\end{equation}
Next differentiate (\ref{6.7}) to $t_n$, then
\[
Res_{z=0}\,
(\frac{1}{z}\left (
\partial_n (P(t,\Lambda))+P(t,\Lambda)\Lambda^n
\right )
e^{\sum t_i\Lambda^i}\epsilon\chi(z)
\Psi^*(s,z)^T)_-=0.
\]
Putting $s=t$ we deduce from this, using (\ref{6.8}) and the Lemma
\ref{l6.1}
that
\[
\partial_n
(P(t,\Lambda))P(t,\Lambda)^{-1}=-(P(t,\Lambda)\Lambda^nP(t,\Lambda)^{-1})_-,
\]
which is the Sato--Wilson equation for the 1-Toda lattice hierarchy.
Differentiate $L(t,\Lambda)=P(t,\Lambda)\Lambda P(t,\Lambda)^{-1}$ to
$t_n$,
then using this Sato--Wilson equation one shows easily that $L$
satisfies (\ref{6.1}).\hfill$\square$

\noindent
As a consequence of the proof of Proposition \ref{p6.1} one has the
Sato--Wilson equation
\begin{equation}
\partial_n
(P(t,\Lambda))P(t,\Lambda)^{-1}=-(P(t,\Lambda)\Lambda^nP(t,\Lambda)^{-1})_-.
\end{equation}

Now return to the Segal--Wilson setting, then part (b) of the above
proposition tells us that the planes $W_n\in Gr^{(n)}(H)$ related to
$\psi_n$
form an infinite flag, in fact one has two infinite flags
\begin{equation}
\label{6.9}
\begin{aligned}
\cdots&\subset& W_{n+1} &\subset& W_{n}&\subset& W_{n-1}&\subset&
W_{n-2}&\subset&\cdots,\\
\cdots&\supset& W_{n+1}^\perp &\supset& W_{n}^\perp&\supset&
W_{n-1}^\perp&\supset& W_{n-2}^\perp&\supset&\cdots.\\
\end{aligned}
\end{equation}
So starting with one plane $W\in Gr^{(n)}(H)$, with wavefunction
$\psi_W(t,z)$, adjoint wavefunction
$\psi^*_W(t,z)$ and tau-function $\tau_W$,
we can apply recursively B\"acklund-Darboux transformations and
adjoint B\"acklund-Darboux transformations to construct a whole flag
of spaces. To be more precise:
\begin{proposition}
\label{p6.2}Let $W\in Gr^{(n)}(H)$, with wavefunction
$\psi_W(t,z)$, adjoint wavefunction
$\psi^*_W(t,z)$ and tau-function $\tau_W$ and let
$w_n,w_{n+1},w_{n+2},\ldots$ be an
ordered Hilbert basis of $W$
and let $w_{n-1}, w_{n-2},w_{n-3},\ldots$
be an ordered Hilbert basis of $W^\perp$,
then the spaces
\begin{equation}
W_m=
\begin{cases}
\{ w\in W|<w|w_k>=0\mbox{ for }k<m\}&\mbox{for } m>n,\\
W&\mbox{if }n=m,\\
W\oplus \sum_{k=m}^{n-1}\mathbb{C}w_k&\mbox{for } m<n
\end{cases}
\end{equation}
and spaces $W_m^\perp$ form infinite flags of the form (\ref{6.9}).
The corresponding wavefunctions $\psi_m(t,z)$, adjoint wavefunction
$\psi^*_m(t,z)$ and tau-function $\tau_m$ are equal to (cf. (\ref{5.1}))
\begin{equation}
\psi_m=
\begin{cases}{\cal W}(\psi_W;w_{n},\dots ,w_{m-1};\psi^*_W;\emptyset)^{-1}
{\cal W}(\psi_W;w_{n},\dots
,w_{m-1};\psi^*_W;\emptyset;\psi_W)&\mbox{for } m>n,\\
\psi_W&\mbox{if }n=m,\\
{\cal W}(\psi_W;\emptyset;w_{m}^\dagger ,\ldots,w_{n-1}^\dagger)^{-1}
{\cal
W}(\psi_W;\emptyset;\psi^*_W;w_{m}^\dagger,\ldots,w_{n-1}^\dagger;
\psi_W)&\mbox{for } m<n
\end{cases}
\end{equation}
and
\begin{equation}
\psi_m^*=
\begin{cases}(-)^{n+m+1}{\cal
W}(\psi_W^*;\emptyset;\psi_W;w_n,\ldots,w_{m-1})^{-1}{\cal
W}(\psi_W^*;\emptyset;\psi_W;w_n,\ldots,w_{m-1};\psi_W^*)&\mbox{for } m>n,\\
\psi_W^*&\mbox{if }n=m,\\
(-)^{n+m+1}{\cal
W}(\psi_W^*;w_{m}^\dagger,\ldots,w_{n-1}^\dagger;\psi_W;\emptyset)^{-1}{\cal
W}(\psi_W^*;w_{m}^\dagger,\ldots,w_{n-1}^\dagger;\psi_W;\emptyset;\psi_W^*)
&\mbox{for } m<n
\end{cases}
\end{equation}
\begin{equation}
\tau_m=
\begin{cases}{\cal W}(\psi_W;w_n,\ldots ,w_{m-1};
\psi^*_W;\emptyset)\tau_W&\mbox{for } m>n,\\
\tau_W&\mbox{if }n=m,\\
{\cal W}(\psi_W;\emptyset;
\psi^*_W;w_{m}^\dagger,\ldots,w_{n-1}^\dagger)\tau_W
&\mbox{for } m<n,
\end{cases}
\end{equation}
Moreover the vector
$\Psi(t,z)=(\epsilon_n\psi_n(t,z))_{n\in\mathbb{Z}}$ is a wavevector and
$\Psi^*(t,z)=(\epsilon_n^{-1}z\psi_{n+1}^*(t,z))_{n\in\mathbb{Z}}$ is its
adjoint wavevector of the 1-Toda lattice hierarchy.
\end{proposition}
The proof of this proposition is straightforward.

\section{Some Applications}

In all the applications that we consider in this section, we start
with $W=W_0=H_+$, hence
$\tau_0=1$, $\psi_0(t,z)=e^{\sum_i t_iz^i}$ and
$\psi_0^*(t,z)=e^{-\sum_i t_iz^i}$.
Let $w_j$, $0\le j\le n-1$, be the following elements of $H_+$:
\begin{equation}
\label{8.1}
w_j=\sum_{i=0}^\infty \overline{\xi}_{ij}z^i,\quad\mbox{with }
\sum_{i=0}^\infty |{\xi}_{ij}|^2<\infty.
\end{equation}
Then for $e^{\sum_i t_iz^i}=\sum_{\ell\in\mathbb{Z}}p_\ell(t)z^\ell$ one
finds the following tau-functions:
\begin{equation}
\label{8.2}
\begin{aligned}
\tau_n&={\cal W}(\psi_0;w_0,\ldots ,w_{n-1};
\psi^*_0;\emptyset )\cr
&=\det(<\frac{\partial^ie^{\sum_k t_kz^k}}{\partial
x^i}|\sum_{\ell=0}^\infty
\overline{\xi}_{\ell j}z^\ell>)_{0\le i,j< n}\cr
&=\det(\sum_{\ell=0}^\infty p_{\ell-i}\xi_{\ell j})_{0\le i,j< n}.
\end{aligned}
\end{equation}
In a similar way we also find for
\begin{equation}
\label{8.3}
w_j=\sum_{i=1}^\infty \overline{\xi}_{-i-j}z^{-i}\in H_-,\quad\mbox{with }
\sum_{i=1}^\infty |{\xi}_{-i-j}|^2<\infty,
\end{equation}
that
\begin{equation}
\label{8.4}
\tau_{-n}=\det(\sum_{\ell=1}^\infty p_{\ell-i}\xi_{-\ell -j})_{1\le
i,j\le n}
\end{equation}
is a KP tau-function.

Notice that if we let $n$ go to infinity, (\ref{8.2}) is a
special case of the famous
result of Sato (see e.g. \cite{M}).

By choosing special elements $w_j$, one obtains special KP
tau-functions. For instance
\[
w_j=\sum_{k=0}^{N-1}\overline{\eta}_{kj}\sum_{i=0}^\infty
(\overline{\lambda_k}z)^i
\]
leads to Hirota's $N$ soliton solution (see also \cite{M}):
\[
\tau_n=\sum_{0\le i_0 < i_1 ,\cdots ,i_{n-1}\le
N-1}e^{\sum_{k=0}^{n-1}\sum_{j=1}^\infty
t_j\lambda_{i_k}^j}\Delta(\lambda_{i_0},\ldots\lambda_{i_{n-1}})
\det(\eta_{i_p q})_{0\le p,q< n},
\]
where
$\Delta(\lambda_{i_0},\ldots\lambda_{i_{n-1}})$ is the Vandermonde
determinant.

Now return to the expressions (\ref{8.2}) and (\ref{8.4}). Here we
have shown a well-known result of
Ohta, Satsuma, Takahashi and Tokihiro \cite{OSTT}, viz that if
$f_0,f_1,\ldots,f_{n-1}$ (resp. $f_{-1},f_{-2},\ldots,f_{-n}$) are
functions satisfying:
\[
\begin{aligned}
\frac{\partial f_j}{\partial t_k}&=\frac{\partial^k f_j}{\partial
x^k}\qquad\qquad\mbox{for }j\ge 0\ \mbox{and}\\
\frac{\partial f_j}{\partial t_k}&=(-)^{k-1}\frac{\partial^k f_j}{\partial
x^k}\quad\mbox{for }j< 0,
\end{aligned}
\]
then the Wronskian determinants
\begin{equation}
\label{8.5}
{\cal W}(f_0,f_1,\ldots,f_{n-1}),\quad (\mbox{resp. }
{\cal W}(f_{-1},f_{-2},\ldots,f_{-n})
\end{equation}
are tau-functions of the KP hierarchy.

In the article \cite{OS}, Oevel and Strampp determine when the Wronskian
solutions of the form (\ref{8.5}) belong to the so-called
$m$-vector $k$-constrained KP hierarchy. This is a reduction of the
KP hierarchy, generalizing the Gelfand-Dickey hierarchies. Here one
assumes that the KP Lax operator $L$ satisfies
\[
L^k=(L^k)_++\sum_{j=1}^m q_j(t)\partial^{-1}r_j(t),
\]
with $q_j(t)$ eigenfunctions and $r_j(t)$ adjoint eigenfunctions of
$L$.
The authors gave in \cite{HvdL1},
\cite{HvdL2} a geometrical interpretation of this $m$-vector
$k$-constrained KP hierarchy in the Segal-Wilson setting. They showed
that $L=L_W$ belongs to this $m$-vector $k$-constrained KP hierarchy
if and only if the corresponding $W\in Gr(H)$ contains a subspace $W'$ of
codimension $m$ such that $z^kW'\subset W$.
We will now use this result to prove geometrically the following
result of Oevel and Strampp \cite{OS}:

\begin{proposition}
\label{P8.1} the tau functions ${\cal W}(f_0,f_1,\ldots,f_{n-1})$ (resp.
${\cal W}(f_{-1},f_{-2},\ldots,f_{-n})$) belong to the $m$-vector
$k$-constrained KP hierarchy if and only if
\[
\begin{aligned}
{\cal W}(f_0,f_1,\ldots,f_{n-1},\frac{\partial^k f_{i_1}}{\partial
x^k},\ldots,\frac{\partial^k f_{i_{m+1}}}{\partial
x^k})&=0,\\ \mbox{(resp. }
{\cal W}(f_{-1},f_{-2},\ldots,f_{-n}))\frac{\partial^k f_{i_1}}{\partial
x^k},\ldots,\frac{\partial^k f_{i_{m+1}}}{\partial
x^k})&=0),
\end{aligned}
\]
for all $(m+1)$-tuples $(f_{i_1},\ldots ,f_{i_{m+1}})$ with indices
$0\le i_1<i_2<\ldots<i_{m+1}\le n-1$ (resp. $-n\le
i_1<i_2<\ldots<i_{m+1}\le -1$).
\end{proposition}
{\bf Proof.}
First consider the positive case, viz $n\ge 0$.
Write $f_j=<e^{\sum_kt_kz^k}|w_j>$ with $w_j$ as in (\ref{8.1}). Then
\[
\tau_n={\cal W}(f_0,f_1,\ldots,f_{n-1})={\cal W}(\psi_0;w_0,\ldots ,w_{n-1};
\psi^*_0;\emptyset ).
\]
If this is a solution of
the $m$-vector
$k$-constrained KP hierarchy, then the space
\[
W_n=\{ v\in H_+|<v|w_i>=0\quad\mbox{for }0\le i\le n-1\}
\]
has a subspace $W'$ of codimension $m$ such that $z^kW'\subset W$.
In other words, one has that for all $w\in W'$  and $0\le j\le n-1$
both $<w|w_j>=0$ and
$<z^kw|w_j>=<w|z^{-k}w_j>=0$.
Which means that
\[
W'\subset \{ v\in H_+|<v|w_i>=<v|z^{-k}w_i>=0\quad\mbox{for }0\le i\le
n-1\}
\subset W.
\]
or equivalently that
\[
\dim <w_0,w_1,\ldots,w_{n-1},
(z^{-k}w_0)_+,(z^{-k}w_1)_+,\ldots,(z^{-k}w_{n-1})_+>
\le n+m,
\]
where $u_+=\sum_{k=0}^\infty u_k z^k$ if $u=\sum_{k\in{\mathbb Z}} u_k
z^k$.
In other words
\[
\begin{aligned}
{\cal W}(f_0,f_1,\ldots,f_{n-1},\frac{\partial^k f_{i_1}}{\partial
x^k},\ldots,\frac{\partial^k f_{i_{m+1}}}{\partial
x^k})&=\\
{\cal W}(\psi_0;w_0,\ldots ,w_{n-1},z^{-k}w_{i_1},\ldots ,z^{-k}w_{i_{m+1}} ;
\psi^*_0;\emptyset )&=0,
\end{aligned}
\]
for all $0\le i_1<i_2<\ldots<i_{m+1}\le n-1$.

In the negative case, one  writes
$f_j=<e^{-\sum_kt_kz^k}|w_j>$ with $w_j$ as in
(\ref{8.3}). One then has to notice that  the $m$-vector
$k$-constrained KP hierarchy means that  $W$ is a subspace  of
codimension $m$ of $W''$ such that $z^kW\subset W''$ or
equivalently $z^{-k}(W'')^\perp\subset W^\perp$ and one can finish
the proof along the same lines as in the positive case.
\hfill$\square$

\vskip 10pt

The next application generates certain orthogonal polynomials. It is
related to Matrix models, see e.g \cite{AvM3}, \cite{HH} and references
therein.

Let $(a,b)$ be an interval of $\mathbb{R}$ and $w(y)$ a weight
function which is non-negative. With these data we
associate the inner product
\begin{equation}
\label{10.1}
(f_1|f_2)=\int_a^b f_1(y)f_2(y)w(y)dy,
\end{equation}
which is well defined for all functions $f$ for which
$\sqrt wf$ is quadratically integrable on $(a,b)$.

The numbers
\begin{equation}
\label{10.2}
c_n=\int_a^b y^nw(y)dy,
\end{equation}
are called the {\it moments} of the weight function $w$.
It is well-known (see e.g. \cite{B}) that the
polynomials
\begin{equation}
\label{10.2a}
p_n(y)=
{{
\det
\left (
\begin{array}{cccc}
c_0&c_1&\cdots&c_n\\
c_1&c_2&\cdots&c_{n+1}\\
\vdots&\vdots&\dots&\vdots\\
c_{n-1}&c_n&\cdots&c_{2n-1}\\
1&y&\cdots&y^n
\end{array}
\right )
}
   \over
{\det (c_{i+j})_{0\le i,j\le n-1}}
}
\end{equation}
form an orthogonal set, i.e.,
\[
(p_m|p_n)=\delta_{mn}
{{
\det(c_{i+j})_{0\le i,j\le n}}
\over
{\det(c_{i+j})_{0\le i,j\le n-1}}
}.
\]
Now consider the generating series of the complex conjugates of the moments
\begin{equation}
\label{10.3}
f(z)=\sum_{k=0}^\infty \overline {c}_kz^k.
\end{equation}
Since the moments of the weight function $w$ satisfy the inequality
\[
|c_n|\le \max (|a|,|b|)^nM,
\]
with $M$ a constant, the series $f(z)$ converges on a neighborhood of
zero. Hence there is a $\lambda_0>0$ such that
$\sum_{n\ge 0}|c_n|^2\lambda_0^{2n}<\infty$. For all $\lambda$ with
$0<\lambda\le \lambda_0$ we define the element $f_i^{(\lambda)}(z)\in
H$ by
\begin{equation}
\label{10.11h}
f_i^{(\lambda)}(z)=\sum_{n=0}^\infty
\overline{c}_n\lambda^nz^{n-i},\quad\text{for }i=0,1,2,\ldots
\end{equation}
Now we consider the following chain of subspaces of $H_+$:
\[
W_0^{(\lambda)}=H_+\quad\text{and }
W_n^{(\lambda)}=\{ w\in H_+| <w|f_i^{(\lambda)}>=0\ \mbox{for
}i=0,1,\ldots,n-1\}.
\]
Note that one can find for each $k\ge 1$ a polynomial $q_k^{(\lambda)}$
of degree $k$ such that
$<q_k^{(\lambda)}|f>=0$. Since the topological span of the
$\{ q_k^{(\lambda)}|k\ge 1\}$ has codimension one in $H_+$ and the same
holds for $W_1^{(\lambda)}$, we must have
\[
W_1^{(\lambda)}=\text{the closure of Span}\{
q_k^{(\lambda)}|k\ge 1\}.
\]
Now we like to choose the $\{ q_k^{(\lambda)}\} $ such that for all
$m\ge 1$
\[
W_m^{(\lambda)}=\text{the closure of Span}\{
q_k^{(\lambda)}|k\ge m\}.
\]
Then the whole chain of subspaces $\{ W_m^{(\lambda)}|m\ge 1\}$
has been expressed in the sequence of polynomials $\{
q_k^{(\lambda)}|k\ge 1\}$.
Because of the codimension one argument, it is necessary and
sufficient that for all $m\ge 1$ and all $j\ge 0,\ j<m$,
\[
<q_m^{(\lambda)}|f_j^{(\lambda)}>=0.
\]
This condition will tell you how to make an appropriate choice for the
$\{
q_k^{(\lambda)}|k\ge 1\}$. For if $q_m^{(\lambda)}(y)=\sum_{\ell=0}^m
a_{m\ell}y^\ell$, then we have
\begin{equation}
\label{10.12h}
\begin{aligned}
<q_m^{(\lambda)}|f_j^{(\lambda)}>=&<\sum_{\ell=0}^m
a_{m\ell}y^\ell|\sum_{r=0}\overline{c}_r\lambda^rz^{r-j}>\\
=&\sum_{\ell=0}^m
a_{m\ell}c_{\ell+j}\lambda^j\\
=&\sum_{\ell=0}^m
\int_a^b
a_{m\ell}y^{\ell+j}\lambda^jw(y)dy\\
=&\int_a^b q_m^{(\lambda)}(\lambda y)y^jw(y)dy\\
\end{aligned}
\end{equation}
Hence, if we choose $q_m^{(\lambda)}(\lambda y)=p_m(y)$, then due to
$(y^j,p_m(y))=0$ for $j<m$, we get that (\ref{10.12h}) is equal to
zero. Thus (cf. \cite{HH})
\begin{equation}
W_m^{(\lambda)}=\text{the closure of Span}
\{ p_k\left (\frac{z}{\lambda}\right )|k\ge m\}.
\end{equation}

Next we compute the tau-function corresponding to this chain following
(\ref{8.2}). We consider $N>\max (|a|,|b|,1)$ and the KP-flows from
$\Gamma_+(N)$.
In that case we have
\[
\begin{aligned}
<\frac{\partial^p e^{\sum_k t_kz^k}}{\partial x^p}
|f_q^{(\lambda)}(z)>=&<z^{p+q}e^{\sum_k t_kz^k}|f_0^{(\lambda)}(z)>\\
=&\int_a^b (\lambda y)^{p+q}e^{\sum_k t_k(\lambda y)^k}w(y)dy\\
=&\lambda^{p+q}c_{p+q}^{(\lambda)}(t),\\
\end{aligned}
\]
where $c_n^{(\lambda)}(t)$ is the $n$-th moment of the $t,\lambda$-dependent
inner product
(cf. (\ref{10.2}))
\begin{equation}
\label{10.7}
(f_1(t,z)|f_2(t,z))_t^{(\lambda)}=\int_a^b f_1(t,z)f_2(t,z)
e^{\sum_k t_k(\lambda z)^k}w(z)dz.
\end{equation}
Note that due to the fact that we have taken our flows from
$\Gamma_+(N)$, this is a well-defined inner product for real $t_i$'s,
even if $\lambda=1$.

We find that the to $W_n^{(\lambda)}$ corresponding wave function
$\psi_n^{(\lambda)}(t,z)=Q_n^{(\lambda)}(t,z)e^{\sum_k t_kz^k}$
is equal to
$BD(W_n^{(\lambda)},H_+)e^{\sum_k t_kz^k}$. Hence,
\begin{equation}
\label{10.8}
\begin{aligned}
Q_n^{(\lambda)}&(t,z)=
{{
\det
\left (
\begin{array}{cccc}
c_0^{(\lambda)}(t)&c_1^{(\lambda)}(t)\lambda
&\cdots&c_n^{(\lambda)}(t)\lambda^n\\
c_1^{(\lambda)}(t)\lambda&c_2^{(\lambda)}(t)
\lambda^2&\cdots&c_{n+1}^{(\lambda)}(t)
\lambda^{n+1}\\
\vdots&\vdots&\dots&\vdots\\
c_{n-1}^{(\lambda)}(t)\lambda^{n-1}
&c_n^{(\lambda)}(t)\lambda^n&\cdots&c_{2n-1}^{(\lambda)}(t)\lambda^{2n-1}\\
1&z&\cdots&z^n\\
\end{array}
\right )
}
   \over
{\det (c_{i+j}^{(\lambda)}(t)\lambda^{i+j})_{0\le i,j\le n-1}
}}\\
=&
{{
\det\left (
\left (
\begin{array}{cccc}
1&&0&\\
&\ddots&&\\
0&&\lambda^{n-1}&\\
&&&1\\
\end{array}\right )
\left (
\begin{array}{cccc}
c_0^{(\lambda)}(t)&c_1^{(\lambda)}(t)&\cdots&c_n^{(\lambda)}(t)\\
c_1^{(\lambda)}(t)&c_2^{(\lambda)}(t)&\cdots&c_{n+1}^{(\lambda)}(t)\\
\vdots&\vdots&\dots&\vdots\\
c_{n-1}^{(\lambda)}(t)&c_n^{(\lambda)}(t)&\cdots&c_{2n-1}^{(\lambda)}(t)\\
1&\frac{z}{\lambda}&\cdots&\left (\frac{z}{\lambda}\right )^n\\
\end{array}
\right )
\left (
\begin{array}{ccc}
1&0&\\
&\ddots&\\
0&&\lambda^n\\
\end{array}\right )\right )
}
   \over
{\lambda^{n(n-1)}\det (c_{i+j}^{(\lambda)}(t))_{0\le i,j\le n-1}
}}\\
=&\lambda^nP_n^{(\lambda)}(t,\frac{z}{\lambda}),\\
\end{aligned}
\end{equation}
where
\begin{equation}
\label{10.p}
P_n^{(\lambda)}(t,z)={{
\det
\left (
\begin{array}{cccc}
c_0^{(\lambda)}(t)&c_1^{(\lambda)}(t) &\cdots&c_n^{(\lambda)}(t)\\
c_1^{(\lambda)}(t)&c_2^{(\lambda)}(t)&\cdots&c_{n+1}^{(\lambda)}(t)\\
\vdots&\vdots&\dots&\vdots\\
c_{n-1}^{(\lambda)}(t)&c_n^{(\lambda)}(t)&\cdots&c_{2n-1}^{(\lambda)}(t)\\
1&z&\cdots&z^n\\
\end{array}
\right )
}
   \over
{\det (c_{i+j}^{(\lambda)}(t))_{0\le i,j\le n-1}
}}.
\end{equation}
These $P_n^{(\lambda)}(t,z)$ form the orthogonal polynomials with respect to
the
$t,\lambda$-dependent inner product $(\cdot|\cdot)_t^{(\lambda)}$, defined by
(\ref{10.7}).
Moreover, the $\psi_n^{(\lambda)}$'s for $n=0,1,2,\ldots$
form a set of $1$-Toda wavefunctions and the $\tau_n^{(\lambda)}$'s
or $\sigma_n^{(\lambda)}$'s
where
\begin{equation}
\label{10.9}
\begin{aligned}
\tau_n^{(\lambda)}(t)&=\det (c_{i+j}^{(\lambda)}(t)\lambda^{i+j})_{0\le i,j\le
n-1}
=\lambda^{n(n-1)}\sigma_n^{(\lambda)}(t),\\
\sigma_n^{(\lambda)}(t)&=\det (c_{i+j}^{(\lambda)}(t))_{0\le i,j\le n-1},\\
\end{aligned}
\end{equation}
form a set of $1$-Toda tau-functions.

For all $w_n\in W_n^{(\lambda)}$ we have that $<w_n|f_i^{(\lambda)}>=0$
for  $i=0,1,\ldots,n-1$, then
$<zw_n|f_i^{(\lambda)}>=<w_n|z^{-1}f_i^{(\lambda)}>=<w_n|f_{i+1}^{(\lambda)}>$
is zero for $i=0,1,\ldots,n-2$.
Hence $zW_n^{(\lambda)}\subset W_{n-1}^{(\lambda)}$ and we deduce
that $\psi_n^{(\lambda)}(t,z)$ belongs to the $1$-vector
$1$-constrained KP hierarchy, i.e.,
\[
L_{W_n^{(\lambda)}}=\partial +q_n^{(\lambda)}\partial^{-1}r_n^{(\lambda)},
\]
with $q_n^{(\lambda)}$ and $r_n^{(\lambda)}$ eigenfunctions, respectively
adjoint eigenfunctions of $L_{W_n}$.
{}From the construction in Section 3 of \cite{HvdL1},
it is easy to determine the $q_n^{(\lambda)}$ and $r_n^{(\lambda)}$'s,
in fact
\[
L_{W_n^{(\lambda)}}=\partial
+g_n\frac{\tau_{n+1}^{(\lambda)}}{\tau_n^{(\lambda)}}
\partial^{-1}
\frac{\tau_{n-1}^{(\lambda)}}{\tau_n^{(\lambda)}},
\]
with $g_n\in\mathbb{C}^\times$.
Hence
\[
\begin{aligned}
z\psi_n^{(\lambda)}&=
\frac{\partial \psi_n^{(\lambda)}}{\partial x}+g_n
\frac{\tau_{n+1}^{(\lambda)}}{\tau_n^{(\lambda)}}
\partial^{-1}
\frac{\tau_{n-1}^{(\lambda)}}{\tau_n^{(\lambda)}}(\psi_n^{(\lambda)})\\
&=\frac{\partial \psi_n^{(\lambda)}}{\partial x}+g_n
\frac{\tau_{n+1}^{(\lambda)}}{\tau_n^{(\lambda)}}
\partial^{-1}
\frac{\tau_{n-1}^{(\lambda)}}{\tau_n^{(\lambda)}}
\frac{\tau_{n}^{(\lambda)}}{\tau_{n-1}^{(\lambda)}}
\partial
\frac{\tau_{n-1}^{(\lambda)}}{\tau_n^{(\lambda)}}(\psi_{n-1}^{(\lambda)})\\
&=\frac{\partial \psi_n^{(\lambda)}}{\partial x}+g_n
\frac{\tau_{n+1}^{(\lambda)}\tau_{n-1}^{(\lambda)}}{(\tau_n^{(\lambda)})^2}
\psi_{n-1}^{(\lambda)}.
\end{aligned}
\]
Since also
\[
\psi_{n+1}^{(\lambda)}=\frac{\partial \psi_n^{(\lambda)}}{\partial x}-
\frac{\partial}{\partial x}\left (
\log \frac{\tau_{n+1}^{(\lambda)}}{\tau_n^{(\lambda)}}\right )
\psi_n^{(\lambda)},
\]
we find that
\begin{equation}
\label{psi}
z\psi_n^{(\lambda)}=\psi_{n+1}^{(\lambda)}+\frac{\partial}{\partial x}\left (
\log \frac{\tau_{n+1}^{(\lambda)}}{\tau_n^{(\lambda)}}\right )
\psi_n^{(\lambda)}+g_n
\frac{\tau_{n+1}^{(\lambda)}\tau_{n-1}^{(\lambda)}}{(\tau_n^{(\lambda)})^2}
\psi_{n-1}^{(\lambda)}.
\end{equation}
We can now determine the constant $g_n$. For this purpose we take the
innerproduct $<z\psi_n^{(\lambda)}|f_{n-1}^{(\lambda)}>$. Then from
(\ref{psi}) we deduce that
\[
g_n=\frac{<\psi_n^{(\lambda)}|f_{n}^{(\lambda)}>\tau_n^{(\lambda)}}
{<\psi_{n-1}^{(\lambda)}|f_{n-1}^{(\lambda)}>\tau_{n-1}^{(\lambda)}}
\frac{\tau_n^{(\lambda)}}{\tau_{n+1}^{(\lambda)}}=
\frac{\tau_{n+1}^{(\lambda)}}{\tau_{n}^{(\lambda)}}
\frac{\tau_n^{(\lambda)}}{\tau_{n+1}^{(\lambda)}}=1
\]
Thus we find that
\begin{equation}
\label{P1}
\frac{z}{\lambda}P_n^{(\lambda)}(t,\frac{z}{\lambda})
=P_{n+1}^{(\lambda)}(t,\frac{z}{\lambda})+
\frac{1}{\lambda}\frac{\partial}{\partial x}\left (
\log \frac{\sigma^{(\lambda)}_{n+1}}{\sigma_n^{(\lambda)}}\right )
P_n^{(\lambda)}(t,\frac{z}{\lambda})+\frac{\sigma_{n+1}^{(\lambda)}\sigma
_{n-1}^{(\lambda)}}{(\sigma_n^{(\lambda)})^2}
P_{n-1}^{(\lambda)}(t,\frac{z}{\lambda}).
\end{equation}
This defines the recursion relation for the orthogonal
polynomials $P_n^{(\lambda)}(t,z)$. Notice that the innerproduct
(\ref{10.7}),
the formula's (\ref{10.8}), (\ref{10.p}), (\ref{10.9}) and (\ref{P1})
are also well-defined for $\lambda=1$. Hence
 let
$P_n(t,z)=P_n^{(1)}(t,z)=
Q_n^{(1)}(t,z)$ and $\tau_n=\tau_n^{(1)}=\sigma_n^{(1)}$, then the
orthogonal
polynomials
$P_n(t,z)$, orthogonal with respect to the inner product
$(\cdot|\cdot)_t^{(1)}$,
satisfy the following recursion relation:
\begin{equation}
zP_n(t,z)=P_{n+1}(t,z)+\frac{\partial}{\partial x}\left (
\log \frac{\tau_{n+1}}{\tau_n}\right )
P_n(t,z)+\frac{\tau_{n+1}\tau_{n-1}}{\tau_n^2}
P_{n-1}(t,z).
\end{equation}

\section{Apendix: Alternative Proof of Theorem \ref{t5.1h}}

We want to give an alternative proof for expression of $\tau_{W_q}$ in
Theorem \ref{t5.1h}.
Let $\tau_W$ be the tau-function corresponding to $W\in
Gr^{(\ell)}(H)$,
then $\tau_W$ satisfies the bilinear identity
\begin{equation}
\label{11.1}
Res_{z=0}\tau_W(t-[z^{-1}])
e^{\sum_{i=1}^\infty (t_i-s_i)z^i}\tau_W(s+[z^{-1}])=0.
\end{equation}
Now let the following operator
\[
w^{2\ell}e^{\sum_{i=1}^\infty (t_i+s_i)w^i}
e^{-\sum_{i=1}^\infty (\frac{\partial}{\partial t_i}+ \frac{\partial}{\partial
s_i})
\frac{w^{-i}}{i}}
\]
act on this bilinear identity (\ref{11.1}). Then we obtain after some
calculation that
\[
Res_{z=0}w^{2\ell}\tau_W(t-[z^{-1}]-[w^{-1}])e^{\sum (t_i-\frac{z^i}{i})w^i}
e^{\sum (t_i-s_i)z^i}
w^\ell\tau_W(s-[w^{-1}]+[z^{-1}])e^{\sum (s_i+\frac{z^i}{i})w^i}=0
\]
and hence
\[
\tilde\tau:=\psi_W(t,w)\tau_w(t)=
w^\ell\tau_W(t-[w^{-1}])e^{\sum_{i=1}^\infty t_iw^i}
\]
is again a tau-function, however not a Segal--Wilson tau-function.
Let $\tilde\psi(t,z)$ be the to $\tilde\tau$ corresponding
wavefunction, i.e.,
\[
\tilde\psi(t,z)=z^{\ell+1}\frac{\tau_W(t-[z^{-1}]-[w^{-1}])
e^{\sum_{i=1}^\infty
(t_i-\frac{z^i}{i})w^i}}{\tau_W(t-[w^{-1}])e^{\sum_{i=1}^\infty t_iw^i}}
e^{\sum_{i=1}^\infty t_iz^i},
\]
then
\[
\psi(t,w)\tilde\psi(t,z)=z\psi_W(t-[z^{-1}],w)\psi_W(t,z).
\]
On the other hand $\tilde\psi(t,z)$ can be obtained with
the B\"acklund--Darboux transformation
\[
\partial-\frac{\partial \log(\tilde\tau/\tau)}{\partial x}=
\partial-\psi_W(t,w)^{-1}\frac{\partial \psi_W(t,w)}{\partial x}
\]
from $\psi_W(t,z)$.
So we obtain that
\[
\begin{aligned}
\hbox{Wronskian}(\psi_W(t,z),\psi_W(t,w))&=
\frac{\partial \psi_W(t,z)}{\partial x}\psi_W(t,w)-
\frac{\partial \psi_W(t,w)}{\partial x}\psi_W(t,z)\\
\ &=z\psi_W(t-[z^{-1}],w)\psi_W(t,z).\\
\end{aligned}
\]

Next let $\psi_q(t,z)=q(t)\partial q(t)^{-1}\psi_W(t,z)$ with
$q(t)=<\psi_W(t,w)|s(w)>$,
then
\[
\begin{aligned}
q(t)\psi_q(t,z)=&<\psi_W(t,w)|s(w)>\psi_q(t,z)\\
=&<\psi_W(t,w)|s(w)>\frac{\partial \psi_W(t,z)}{\partial x}-
<\frac{\partial \psi_W(t,w)}{\partial x}|s(w)>\psi_W(t,z)\\
=&<\hbox{Wronskian}(\psi_W(t,z),\psi_W(t,w))|s(w)>\\
=&<z\psi_W(t-[z^{-1}],w)\psi_W(t,z)|s(w)>\\
=&zq(t-[z^{-1}])\psi_W(t,z),
\end{aligned}
\]
and
\begin{equation}
\psi_q(t,z)=z\frac{q(t-[z^{-1}])}{q(t)}\psi_W(t,z).
\end{equation}
Now use formula (\ref{tau}) and one obtains that
\[
\partial_k\log\frac {\tau_{W_q}(t)}{q(t)\tau_W(t)}=0\quad\hbox{for all
}k=1,2,\ldots
\]
Thus we conclude that upto multiplication by a constant
\begin{equation}
\tau_{W_q}(t)=q(t)\tau_W(t).
\end{equation}
The formula $\tau_{W_r}(t)=r(t)\tau_W(t)$ can be obtained in a similar matter.

\end{document}